\documentclass[pdflatex,sn-mathphys-num %,sn-nature
,referee]{sn-jnl}% Math and Physical Sciences Numbered Reference Style 
\usepackage{graphicx}%
\usepackage{multirow}%
\usepackage{amsmath,amssymb,amsfonts}%
\usepackage{amsthm}%
\usepackage{mathrsfs}%
\usepackage{mathtools}
\usepackage[title]{appendix}%
\usepackage{xcolor}%
\usepackage{textcomp}%
\usepackage{upgreek}%
\usepackage{manyfoot}%
\usepackage{booktabs}%
\usepackage{algorithm}%
\usepackage{algorithmicx}%
\usepackage{algpseudocode}%
\usepackage{listings}%
\usepackage{gensymb}
\usepackage{bm}
\usepackage{subcaption}
\usepackage{siunitx} 
\usepackage{placeins}
\usepackage{lineno}
\DeclarePairedDelimiter{\abs}{\lvert}{\rvert}
\newcommand{\extfig}{Extended Data } % for referring to extended data figures and tables
\newcommand{\SIfig}{Supplementary Information } % for referring to SI figures and tables

\begin{document}

\title[Article Title]{%Emergent Single-Electron Transport and Memristive Effects in Molecular Ensemble Junctions via Sulfur-Gold Interface Reconstruction \\
%/ or / \\
%Instability of the gold-thiol bond in molecular self-assembled monolayers: Coulomb blockade in metal clusters \\/ or / \\

%Migration of thiol-bonded gold atoms into a molecular self-assembled monolayer, forming a cluster exhibiting a Coulomb staircase
Migration of gold atoms into a thiol-bonded molecular self-assembled monolayer, forming a cluster exhibiting a Coulomb staircase
% JJB suggests: Coulomb staircase from clusters formed by migration of gold atoms in thiol-bonded self-assembled molecular monolayers
% \\/ or / \\Single-Electron Effects Caused by Cluster Formation from Thiol-Gold Bonds in Molecular Monolayers
}

% CF: Instability of the gold-thiol bond in molecular self-assembled monolayers: Coulomb blockade in metal clusters
%Photo-Induced Charge States Enable Memristive NDR in Molecular Junctions
% Photo-Induced Long-Lived Charge States and Coulomb Blockade in Solid-State Memristive Molecular Junctions with NDR
%Photo-Induced Long-Lived Charge States Enable Negative Differential and Memristive Coulomb Blockade Transport in Solid-State Molecular Junctions
%Coulomb Blockade and Memristive Single-Electron Transport via Sulfur-Metal Interface Restructuring in Molecular Junctions

%%=============================================================%%
%% GivenName	-> \fnm{Joergen W.}
%% Particle	-> \spfx{van der} -> surname prefix
%% FamilyName	-> \sur{Ploeg}
%% Suffix	-> \sfx{IV}
%% \author*[1,2]{\fnm{Joergen W.} \spfx{van der} \sur{Ploeg} 
%%  \sfx{IV}}\email{iauthor@gmail.com}
%%=============================================================%%

\author*[1]{\fnm{Bingxin} \sur{Li}}\email{bl497@cantab.ac.uk}
\equalcont{These authors contributed equally to this work.}

\author*[1]{\fnm{Shanglong} \sur{Ning}}\email{sn538@cantab.ac.uk}
\equalcont{These authors contributed equally to this work.}

\author*[1]{\fnm{Chunyang} \sur{Miao}}\email{cm2109@cam.ac.uk}
\equalcont{These authors contributed equally to this work.}

\author[1]{\fnm{Chenyang} \sur{Guo}}%\email{cg742@cantab.ac.uk}

\author[3]{\fnm{Gyu Don} \sur{Kong}}% yoyokongm@naver.com

\author[1]{\fnm{Xintai} \sur{Wang}}%\email{wangxintai1984@163.com}

\author[1]{\fnm{Victor I.} \sur{Coldea}}%\email{vic21@cam.ac.uk}

\author[1]{\fnm{Yuqiao} \sur{Li}}%\email{yl2007@cam.ac.uk}
\author[2]{\fnm{Sam} \sur{Harley}}%\email{s.harley1@lancaster.ac.uk} ORCID: 0009-0004-9411-3504
\author[2]{\fnm{Oleg V.} \sur{Kolosov}}%ORCID: 0000-0003-3278-9643  o.kolosov@lancaster.ac.uk
\author[2]{\fnm{James} \sur{Newson}}%\email{j.newson@lancaster.ac.uk}
\author[2]{\fnm{Sam P.} \sur{Jarvis}}%\email{samuel.jarvis@lancaster.ac.uk}
\author[2]{\fnm{Ben J.} \sur{Robinson}}%\email{b.j.robinson@lancaster.ac.uk} ORCID: 0000-0001-8676-6469
\author[2]{\fnm{Mohammed} \sur{Alzanbaqi}}% m.alzanbaqi@lancaster.ac.uk
\author[2]{\fnm{Ali} \sur{Ismael}} % k.ismael@lancaster.ac.uk
\author[2]{\fnm{Colin J.} \sur{Lambert}} % c.lambert@lancaster.ac.uk

\author[3]{\fnm{Hyo Jae} \sur{Yoon}}%\email{hyoon@korea.ac.kr}

\author[1]{\fnm{Jeremy J.} \sur{Baumberg}}%\email{jjb12@cam.ac.uk}

\author*[1]{\fnm{Christopher J. B.} \sur{Ford}}\email{cjbf@cam.ac.uk}

\affil*[1]{\orgdiv{Cavendish Laboratory}, \orgname{University of Cambridge}, \orgaddress{\street{J J Thomson Avenue}, \city{Cambridge}, \postcode{CB3 0US}, \country{United Kingdom}}}

%\affil[2]{\orgdiv{Blackett Laboratory}, \orgname{Imperial College London}, \orgaddress{\street{South Kensington Campus}, \city{London}, \postcode{SW7 2AZ}, \country{United Kingdom}}}
\affil[2]{\orgdiv{Physics Department}, \orgname{Lancaster University}, \orgaddress{\city{Lancaster}, \postcode{LA1 4YB}, \country{United Kingdom}}}

\affil[3]{\orgdiv{Department of Chemistry}, \orgname{Korea University}, \orgaddress{\city{Seoul}, \postcode{02841}, \country{Korea}}}

%\affil[4]{\orgdiv{Nanophotonics Centre, Cavendish Laboratory}, \orgname{University of Cambridge}, \orgaddress{\street{J J Thomson Avenue}, \city{Cambridge}, \postcode{CB3 0US}, \country{United Kingdom}}}

\abstract{Thiol-based self-assembled monolayers (SAMs) on gold surfaces are one of the fundamental building blocks of molecular electronics. The strong chemical affinity of the gold and sulfur (Au-S) enables the formation of close-packed SAMs, but it also has recently been found to create a dynamic interface where surface reconstruction can occur under illumination, even with ambient light. This reconstruction may facilitate migration of gold atoms, potentially leading to \textit{in-situ} formation of gold clusters. However, research on this mechanism often centers on Au(111) crystalline surfaces and flicker-noise measurements. Electron transport in ensembles of molecules in lithographically defined junctions has remained largely unexplored at cryogenic temperatures. In this study, we observe single-electron phenomena characterized by reproducible Coulomb staircases across various long-chain alkanethiol SAMs, which fit the Coulomb-blockade theory of nm-sized metallic nanoparticles. We find no such current steps in samples with amine, rather than thiol, anchors. Additionally, we find that by adding a bipyridyl functional group, these phenomena can be harnessed for memristive switching and negative differential resistance. 
These findings indicate that the generally observed lack of reliability and reproducibility of molecular devices may be alleviated by using amine anchors instead of thiols to avoid nanoparticle effects. Conversely, the spontaneous formation of the nanoparticles could potentially be controlled and used to achieve useful functionalities, offering new pathways for designing multifunctional nanoelectronic components.
%These findings have dual implications: researchers seeking predictable molecular conductance may benefit from using amine anchors to avoid nanoparticle effects and hence increase SAM reliability and the reproducibility of molecular devices, while the spontaneously formed nanoparticles themselves could be deliberately used to achieve useful functionalities, offering new pathways for designing multifunctional nanoelectronic components.
}

%%================================%%
%% Sample for structured abstract %%
%%================================%%


%%\pacs[JEL Classification]{D8, H51}

%%\pacs[MSC Classification]{35A01, 65L10, 65L12, 65L20, 65L70}
\large
\maketitle

\section{Introduction}\label{sec1}

Molecular-scale devices have inspired a vision of smaller, faster, and more energy-efficient electronic circuits that push beyond the limits of conventional semiconductor technology \cite{chenMolecularSupramolecularElectronics2021, ratnerBriefHistoryMolecular2013, hushOverviewFirstHalfCentury2003}. Unlike traditional integrated-circuit electronics, which manipulate electron flow through doped semiconductor regions with fixed physical configurations, molecular electronics flexibly makes use of molecules' unique physical and chemical properties, such as quantum interference, tailored electronic states, and tunability of functional groups, to construct devices displaying unconventional electrical and optical characteristics \cite{jensenQuantumComputingMolecular2022,gorgonReversibleSpinopticalInterface2023}. In recent years, significant progress has been made in the fabrication, characterization, and manipulation of molecular junctions. Techniques such as scanning tunneling microscopy (STM) \cite{binnigSurfaceStudiesScanning1982}, mechanically controllable break junctions (MCBJs) \cite{reedConductanceMolecularJunction1997}, lateral junctions formed using lithography with the help of large nanocrystals \cite{dadoshMeasurementConductanceSingle2005}, and graphene single-molecular junctions (SMJs) \cite{zhouDirectObservationSinglemolecule2018} have allowed researchers to probe electronic transport at unprecedented spatial and temporal resolutions. The field is also expanding beyond single-molecule measurements toward complex systems capable of logic operations, memory storage, and energy harvesting at the molecular scale \cite{sharmaLinearSymmetricSelfselecting2024, wangTailoringQuantumTransport2024, kangInterplayFermiLevel2020,parkRectificationMolecularTunneling2021,wang2023orientation}.

Despite these milestones, fabrication of functional, reproducible, scalable and tunable single-molecule junctions at the solid-state device level remains challenging. To translate molecular electronic devices from laboratory investigations to practical, scalable technologies, techniques for making electrical contact to a finite area on top of a self-assembled monolayer (SAM) have been developed \cite{fruhmanHighyieldParallelFabrication2021, nijhuisMechanismRectificationTunneling2010,puebla-hellmannMetallicNanoparticleContacts2018, green160kilobitMolecularElectronic2007, liMolecularEnsembleJunctions2022}. Stochastic behavior in conductance is significantly reduced due to averaging over millions of molecules and the limited variation of binding geometry of the molecules \cite{wangElectrostaticFermiLevel2022}. Using sulfur-terminated molecules (thiols) to anchor the molecules to metals such as gold has been ubiquitous. We have demonstrated that, by contacting a SAM of 1,6 hexanedithiol and colloidal quantum dots (QDs) using single-layer graphene in lithographically defined micron-sized junctions, one can readily observe single-electron transport, manifested as a Coulomb staircase, in these heterojunctions over a broad temperature range from 4\,K up to at least 70\,K \cite{fruhmanHighyieldParallelFabrication2021,ismaelTuningElectricalConductance2024}. In addition, using ionic-liquid gating of various molecular or nanocrystal SAMs through the graphene \cite{wangElectrostaticFermiLevel2022,ismaelTuningElectricalConductance2024}, the position of the energy levels could be systematically tuned, thus enabling refined control over the transport characteristics in these nanoscale systems.

In this paper, we now investigate the emergence of single-electron phenomena in lithographically defined junctions containing a sulfur-anchored molecular SAM sandwiched between graphene and gold electrodes. Fits to each Coulomb staircase indicate that these single-electron effects arise from \textit{metallic} nanoparticles that form in the junction. We have characterized the devices over a broad temperature range, between cryogenic and room temperatures. Instability of the gold surface in the presence of a thiol-anchored SAM, including the presence of gold atoms among the molecules, has been discussed before \cite{vericatSelfassembledMonolayersThiols2010, kautzAlkanethiolAu111SelfAssembled2008}, but it has been hard to quantify the effects on electrical transport when probing only one or a few molecules at a time. Our results here show that often there is just one significant cluster in the whole junction area containing millions of molecules. We have not observed any similar current-voltage steps when we replace thiols with amine (-NH$_2$) anchors, suggesting that the thiol is responsible for dragging Au atoms out of the electrode, after which such Au atoms clump together to form a nanoparticle. In fact, our simulations show that, instead, the amine group traps individual Au atoms that may be present on the irregular Au surface, greatly reducing the chance of cluster formation. Furthermore, we demonstrate that leveraging these single-electron effects with suitable molecules enables the fabrication of devices exhibiting Coulomb blockade, memristive characteristics, and negative differential resistance (NDR). These could potentially perform sensing, memory, and computation in a single, reconfigurable electrical component, offering the opportunity to significantly reduce a processor's power consumption, and hence its carbon footprint.

\section{Results and Discussion}\label{sec2}

\subsection{Device structure and characterization}

Each sample consists of an oxidized silicon chip containing a set of gold electrodes upon which a SAM is grown in selectively etched holes in an aluminum oxide (AlO$_x$) layer. A large piece of single-layer graphene is laid on top as the top electrode, with each junction forming a two-terminal gold--SAM--graphene sandwich, as illustrated in Fig.~\ref{fig: selection of junctions showing staircases or peaks}(a).

% \begin{figure}[htb!]
    
%     \includegraphics[width=\textwidth]{schematics4.pdf}
%     \caption{Device schematics. (a) is a three-dimensional rendered view of a single junction showing a monolayer of molecules assembled on a gold layer inside a hole etched through AlOx layer. The top contact is graphene. (b) is the cross-sectional view of the junction highlighting each individual layer. The arrows illustrates the process of the S-bond pulling out gold naonoparticles from the gold substrate.}
%     \label{fig:microwelldesign}
% \end{figure}

\subsection{Single-electron tunneling}

\begin{figure}[htbp!]
	\centering
    % \begin{subfigure}[b]{\textwidth}
    %     \includegraphics[width=1\textwidth]{thesisFigureOutlook}
    % \end{subfigure}

    % \begin{subfigure}[b]{\textwidth}
    %     \includegraphics[width=1\textwidth]{schematics 5.pdf}
    % \end{subfigure}
    \includegraphics[width=1\textwidth]{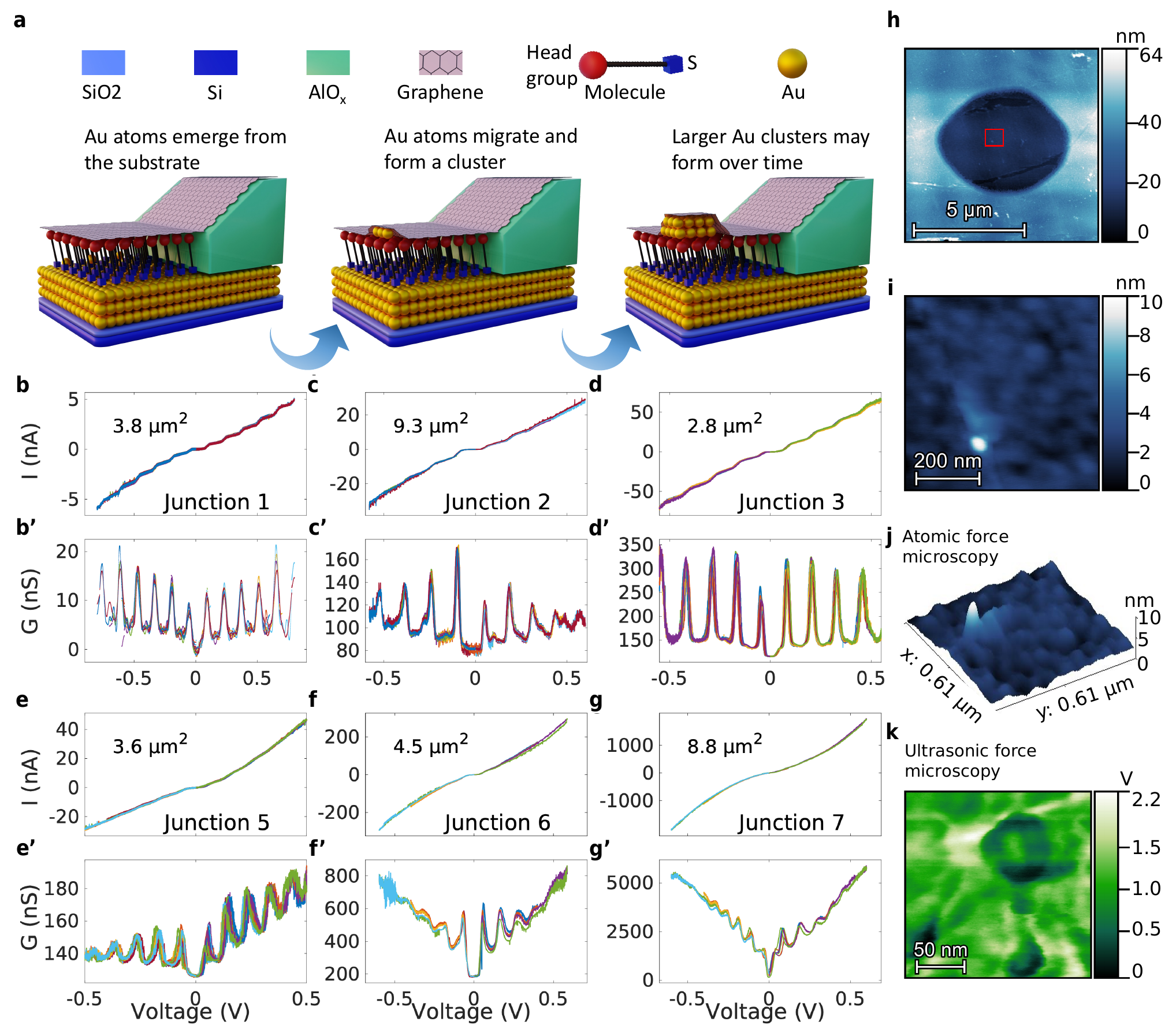}
	\caption{ (a) Cross-sectional schematic renderings of the junction, showing a monolayer of molecules assembled on a gold layer inside a hole etched through an insulating layer. The top contact is graphene. A gold cluster, here shown as a disk beneath the graphene, may form in a pristine monolayer as Au atoms migrate from the surface, as indicated by the arrows between diagrams. (b)-(g) A collection of junctions with various areas from different batches but showing similar pronounced staircases in the $I$--$V$ characteristics (the first and third rows) with corresponding multiple peaks in the conductance ($G$) spectra (the second and fourth rows) at 4.2\,K. Curves with different colors represent consecutive sweeps, showing reproducibility. (h) Height map captured with PeakForce tapping AFM for a microwell containing a thiol-anchored SAM, showing only a few white specks that may be clusters. (i) High-resolution AFM height map of the red box in (h). (j) 3D view of (i), showing a single bump under the graphene. (k) Image of a bump in another microwell junction using ultrasonic force microscopy, showing a dark ring of unsupported graphene and a firm region in the middle of the bump (lighter), which we attribute to a gold cluster less than 15\,nm in diameter and 10\,nm in height, possibly a little larger than one that gives rise to a Coulomb staircase.}
	\label{fig: selection of junctions showing staircases or peaks} 
\end{figure}

A total of 91 junctions from 5 batches of the SAM of 1-hexadecanethiol ($\text{CH}_3(\text{CH}_2)_{15}\text{SH}$, abbreviated to C16S) were measured at 4.2\,K directly after fabrication. Approximately 33\% of the junctions showed clear reproducible staircases in the current $I$ as a function of voltage $V$ and corresponding distinct peaks in the differential conductance $G$.

Fig.~\ref{fig: selection of junctions showing staircases or peaks} presents the $I$--$V$ and corresponding $G$--$V$ curves for a selection of 6 junctions, to capture the variation of the staircase shape between junctions. The junctions in plots (b) and (d) are from different batches, but both show a similar staircase. Eight consecutive pronounced steps can be observed within $\pm0.6$\,V for these two junctions. The steps occur at approximately the same locations with approximately the same spacing. The junctions in (c) and (e) also show staircases, but with irregular step heights. Junctions in (f) and (g) exhibit less-pronounced staircases, but each still shows a regular series of distinct conductance peaks. 

These characteristics in our molecular junctions resemble the signature of single-electron sequential tunneling via a confined state in the molecule. However, such Coulomb-staircase phenomena are not expected in alkanethiol SAM junctions such as C16S, as these saturated alkyl chains typically have a large HOMO-LUMO gap ($\sim$8 \,eV), which forms a high tunnel barrier along the whole molecule. This prevents there being any low-energy state where charge could be localized, as required for Coulomb blockade. 

% \subsection{Theory of Coulomb blockade}

\begin{figure}[htbp!]
    \centering
    \includegraphics[width=1\textwidth]{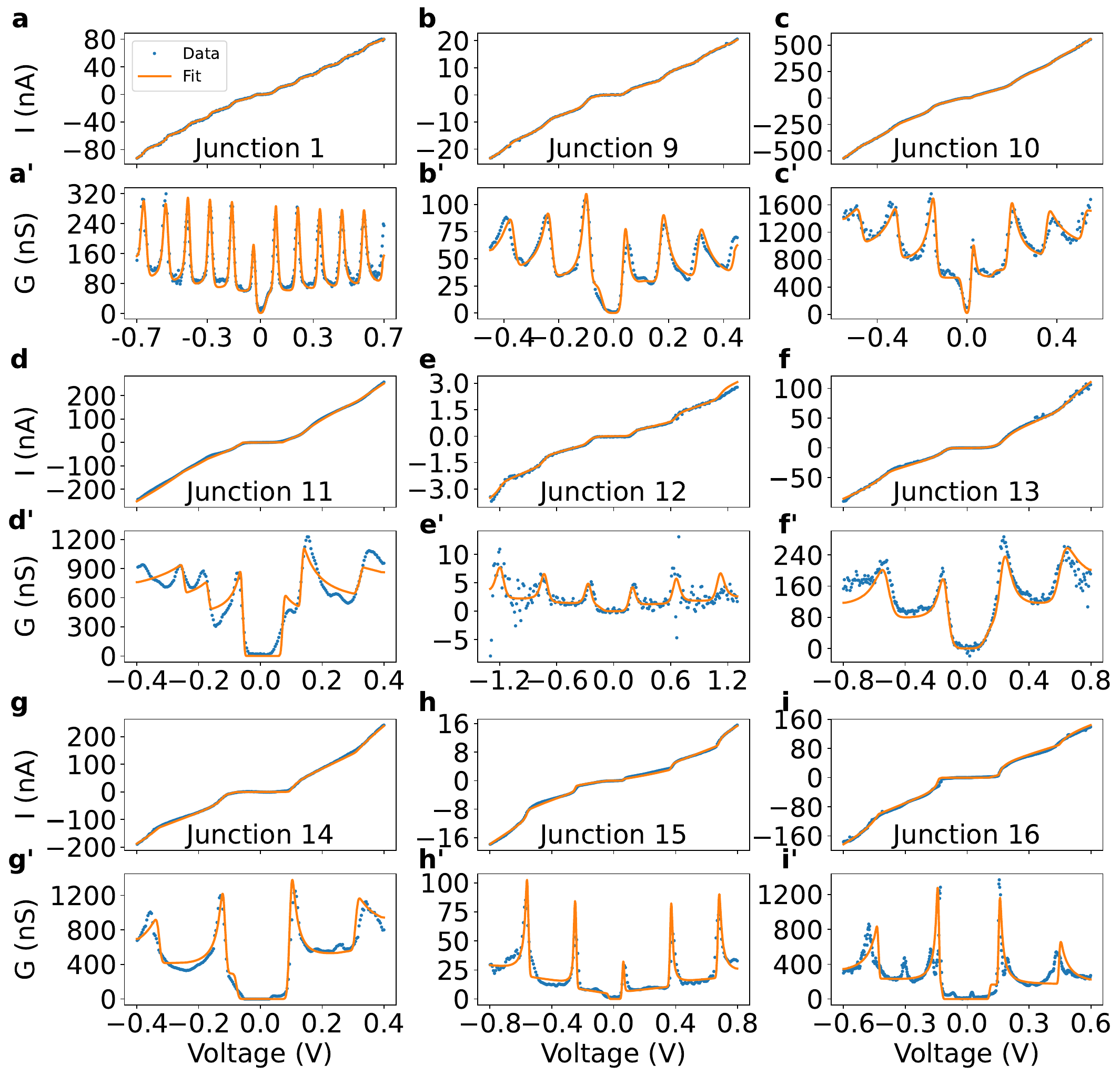}
    \caption{$I$--$V$ and conductance (G) measurements (blue dots) and their corresponding fits (orange curves, see text), for a representative variety of junctions. (a), (b), (g) have a C16S SAM, (c), (h), (i) have HSC11BIPY with C10S2 supporting molecules, (d) has C18S, and (e) and (f) have HSC11BIPY with C8S2 supporting molecules, where 2.2\,nm and 3\,nm gold nanoparticles were deliberately introduced, respectively. Except for (e) and (f), for which the measurements were conducted at 77\,K, the rest of the measurements were carried out at 4.2\,K.}
    \label{fig:ivcurves}
\end{figure}

To investigate the origin of the observed $I$--$V$ staircases, we fabricated additional junctions using alkanethiols of various chain lengths, both shorter and longer than the C16S molecules. We also used 2,2$^\prime$-bipyridyl-terminated n-undecanethiol (HSC11BIPY or BIPY for short), a molecule that features a distinctive functional group whilst maintaining the thiol anchoring (see structure in Fig.\ \ref{fig:background_charge_gating}(a)) \cite{kongInterstitiallyMixedSelfAssembled2021,jangElectricallyStableSelfAssembled2024}. Notably, the $I$--$V$ staircases were consistently observed for alkanethiol chains containing at least 14 carbons, and for the HSC11BIPY SAMs.
Fig.~\ref{fig:ivcurves} shows representative $I$--$V$ curves measured at cryogenic temperatures (4.2\,K or 77\,K) for various SAM compositions. Although most junctions show a clear tunneling-like $I$--$V$ characteristic at room temperature (see \extfig Fig.\ \ref{fig:temperature compare}), at cryogenic temperatures we observe a pronounced series of approximately equally spaced, step-like features in the $I$--$V$ curve, with a region of suppressed conductance around zero bias---hallmarks of Coulomb blockade in a metallic or quantum island (`dot'). It is worth noting that some junctions show similar conductance regardless of the temperature, while others exhibit a significant increase in conductance at cryogenic temperatures. The voltage spacing $\Delta V$ between consecutive current steps or conductance peaks is related to the charging (addition) energy $E_{\rm c} = e^2/C$, where $C$ is the total capacitance of the dot to the surrounding conductors.

The nearly equidistant steps in many $I$--$V$ characteristics, and the corresponding conductance peaks, combined with linear current increases after each step, provide compelling evidence for single-electron tunneling through a metallic island (`dot'), as opposed to resonant tunneling through discrete molecular orbitals~\cite{brunDynamicalCoulombBlockade2012}.  This periodicity in voltage ($\Delta V \approx e/C$) emerges from the discrete charging of a nanoscale capacitor, as described by the theory of classical Coulomb blockade. The observed asymmetry of the conductance peaks---typically sharper on the low-bias side---contrasts sharply with the symmetric Lorentzian peaks expected for coherent tunneling through molecular energy levels. This asymmetry and the corresponding finite slope after the current step arise from the increased number of single-particle states available to tunnel through, as the energy (bias) window increases. 

The series of steps forming a Coulomb staircase results from the asymmetry in the tunnel barriers: when the bias becomes high enough to overcome the charging energy to add an extra electron to the dot, for one bias polarity it is easier to tunnel into the dot than out, so the dot gains an extra electron for most of the time. Each electron there has a chance of tunneling out, so the addition of another gives a step increase in the total tunneling probability and hence an extra current step. Under the opposite bias, there are fewer electrons in the dot, as they can tunnel out easily and be replenished more slowly. The barrier asymmetry may arise from a random offset of the island between the electrodes, or from the difference in tunneling through the anchor group and into the graphene.
%Long-chain molecules such as C16S could provide the requisite tunnel-barrier asymmetry due to its high resistivity.

\begin{figure}[tb!]
    \centering
    \begin{subfigure}[b]{0.48\textwidth}
    \includegraphics[width=\textwidth]{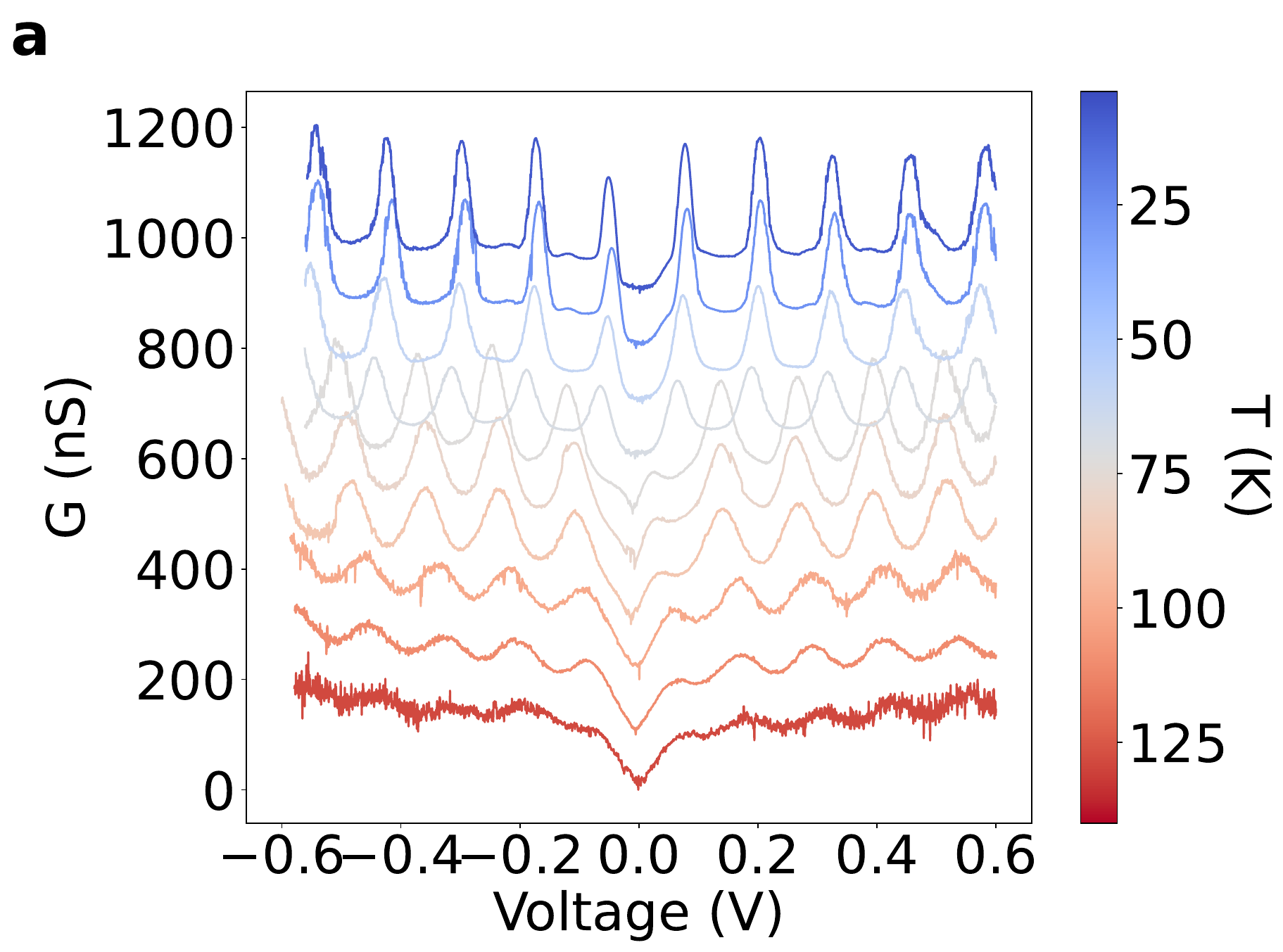}
    \end{subfigure}
    \begin{subfigure}[b]{0.48\textwidth}
    \includegraphics[width=\textwidth]{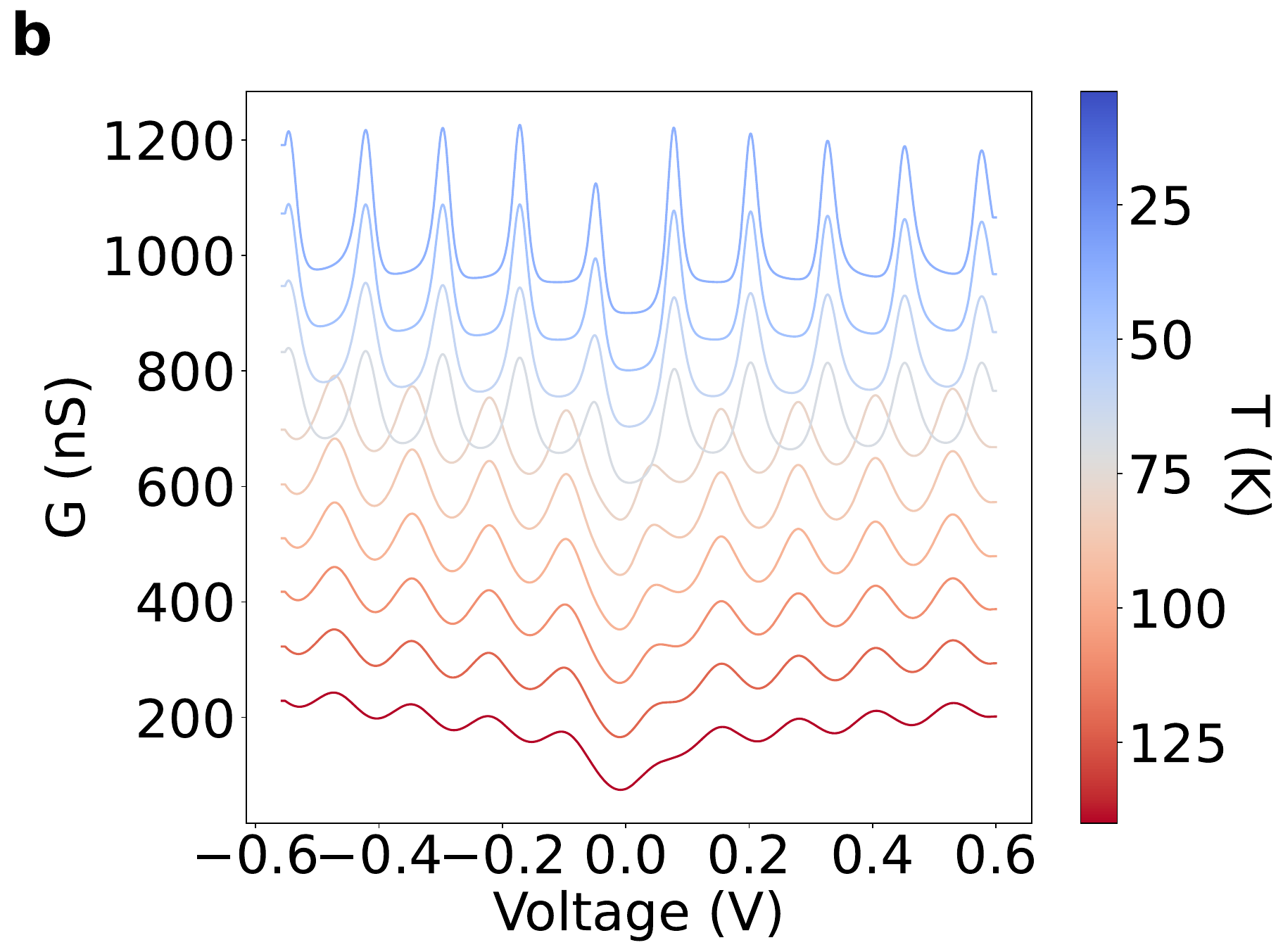}
    \end{subfigure}
    \caption{Temperature-dependence measurements of Junction 1. (a) Experimental data, (b) fits to each of the curves in (a) The color of each curve corresponds to the temperature given by the colorbar to the right of each plot, which includes any other forms of energy broadening. A sudden change occurred in the junction before the fifth trace from the top while warming from low temperature (at around 70\,K). The corresponding main change in the fitting parameters (from the set labeled 1$^\prime$ to that labeled 1$^{\prime \prime}$ in Table \ref{tab:parameters}) is in the background charge $Q_0$, where the values of $Q_0/e$ below and above 70\,K are $0.12\pm0.01$ and $-0.26\pm0.01$, respectively. The other fitting parameters change by less than 20\%.}
    \label{fig:tdpendence}
\end{figure}

Fig.~\ref{fig:tdpendence} presents the temperature-dependent transport characteristics of Junction 1. With increasing temperature, thermal excitations progressively broaden the Coulomb-blockade features as the thermal energy ($k_{\rm B}T$) approaches the charging energy. Single-electron tunneling persists up to at least $\sim$130\,K, beyond which thermal broadening obscures the quantized nature of electron transport, yielding smooth $I$--$V$ characteristics with reduced contrast in conductance measurements.

The presence of robust Coulomb blockade and Coulomb staircases in these junctions indicates the formation of nanometer-sized metallic islands within the molecular layer, as determined by the excellent fits that will be described in the next section. We attribute this to the presence of gold clusters or nanoparticles that aggregate spontaneously during, or after, SAM growth, as thiol-anchored molecules have been shown to harbor single gold atoms and gold clusters \cite{guoExtensivePhotochemicalRestructuring2024, huangEnhancedPhotocurrentElectrically2024, baumbergQuantumPlasmonicsSubAtomThick2023}, perhaps because the gold-thiol bond is strong enough to dislodge gold atoms from the substrate \cite{hakkinen_goldsulfur_2012}, or, as our modeling in section \ref{secModeling} shows, because ubiquitous adatoms are not trapped by thiol groups (whereas they are trapped by amine groups). The movement of Au atoms into a possible cluster configuration within the SAM are sketched in Fig.~\ref{fig: selection of junctions showing staircases or peaks}(a). % These nanoparticles create a double-barrier tunnel-junction configuration, with the molecular SAM serving as tunnel barriers between the nanoparticle and electrodes. 

We have recently shown that strong illumination of SAMs of certain molecules causes progressive growth of very large clusters on top of the SAM \cite{guoExtensivePhotochemicalRestructuring2024}. Our junctions only receive UV illumination at most once, to break up resist on the gold just before assembling the SAM, and the chips are kept in ambient light until they are measured electrically, so some light-induced cluster cannot be ruled out. It is therefore plausible that, even without intentional illumination, a single cluster of tens or hundreds of atoms can form within a few square microns (containing millions of molecules), though it would be almost impossible to spot, given the other irregularities and impurities within such an area. By decreasing the SAM conductivity to below $\sim1\,\upmu$S/$\upmu$m$^2$ using long molecules such as C16S, the staircase current through the cluster becomes noticeable or dominant (the graphene series conductance is 100--1000 times higher than this and is therefore unimportant). 

Using atomic force microscopy (AFM), we studied junctions that had been measured and found to show staircase structure (see \ref{SI_AFM_UFM}). Within each, one or a few bumps were found on the graphene, up to 9\,nm high (see Fig.\ \ref{fig: selection of junctions showing staircases or peaks}(h--k)). Ultrasonic force microscopy (UFM) was then used to map the restoring force of the graphene, which was found to be low except at the center of each bump, implying that the graphene is stretched over some cluster beneath it, and freestanding around it. This is consistent with the existence of gold clusters. These are somewhat larger than the diameters calculated from the capacitances from fitting the Coulomb staircase results. Indeed, junction 11 showed a good staircase, but no detectable bump, whereas the other wells shown here showed less regular behavior, indicating perhaps that the clusters that give good Coulomb blockade are smaller than those that are detectable even with the best AFM and UFM measurements.

This cluster model is further supported by our observation that a higher proportion of junctions display Coulomb staircases when measured three months after their fabrication, suggesting continued nanoparticle formation over time \cite{grysCitrateCoordinationBridging2020}. S--Au bond cleavage may also be caused by exposure to air, as it may reduce the Au--Au rupture force and facilitate some pulling out of gold atoms, but the graphene and photoresist over our junctions limit such air ingress. During junction fabrication lasting days or weeks, we do not currently limit the exposure to visible light, so it is as yet unclear whether some light is needed to promote cluster growth. We have illuminated our HSC11BIPY SAMs for many hours and seen the formation of a few visible clusters (see Supplementary Information, \extfig Figs.~\ref{fig:optics}--\ref{fig:DF vs BF}). However, no change was observed in C16N and C16S SAMs. This does not rule out formation of small clusters, only their further growth into visible islands, which is enhanced by the particular structure of the 4,4$^\prime$-biphenyldithiol (BPD) molecules tested in \cite{guoExtensivePhotochemicalRestructuring2024}.
Likewise, Au–bipyridine coordination in HSC11BIPY may stabilize the single gold atoms and facilitate their aggregation. Moreover, graphene could also facilitate the diffusion of gold atoms, and gold clusters could nucleate once one Au atom becomes trapped at a dangling bond \cite{joudiTwoDimensionalOneAtomThickGold2025}.

Another possible means of gold-cluster formation might come from the top monolayer of gold atoms becoming detached from the gold substrate over a small area, probably again because of the force from the thiol bond \cite{baumbergQuantumPlasmonicsSubAtomThick2023}. The tunneling probability through the anchoring Au atoms around the edge might be so low that the number of electrons on the cluster is well-defined (with resistance to the rest of the gold $>h/e^2$), though this seems unlikely, and it is not clear how stable such a disk of atoms would remain as the forces change while sweeping the bias. 

%By adopting the technique described in \cite{guoExtensivePhotochemicalRestructuring2024}, we observe the formation of gold nanocaps in HSC11BIPY SAM. However, no evident change was observed in C16N and C16S SAM, as shown in \extfig Fig.~\ref{fig:BF illumination} and Fig.~\ref{fig:DF vs BF}. 

To test whether the gold-thiol bond is important for this phenomenon, we replaced C16S with C16N, where the thiol ($\mathrm{-SH}$) is replaced by an amine group ($\mathrm{-NH_2}$), and indeed we observe no Coulomb blockade for such SAMs. We also find that C16N devices show a much narrower current-density distribution (Fig. \ref{fig:background_charge_gating}(g)), suggesting that C16N forms better-quality SAMs with less random variation. 

High-resolution X-ray photoemission spectroscopy (XPS) was used to check for contamination from the graphene capping process and also SAM formation. \ref{fig:SXPS1} shows XPS in the O 1s, C 1s, and Cu 2p regions, showing results typical of graphene on PMMA. Despite using exceptionally long acquisition times, and attempts to increase surface sensitivity using grazing-angle XPS, no Cu 2p was identified. Similarly, \ref{fig:SXPS2} shows XPS on a C16N SAM. Here we observe a very weak Cu 2p signal, corresponding to an atomic percentage of $<$0.15\%, which we attribute to solvent contamination. No other contamination signals were identified. The low Cu signal, combined with the lack of Coulomb staircase in the C16N SAMs, which used the same solvents, rules out the possibility that such impurities give rise to the Coulomb staircases.

A subsidiary peak is often observed after each principal Coulomb-blockade conductance peak, separated from it by about 40--100\,mV. They are observed in repeated voltage sweeps, across multiple devices, and under varying temperature conditions, pointing to intrinsic quantum-mechanical behavior as opposed to random background fluctuations. The energy spacing is therefore of order 20--50\,meV, if roughly half the voltage drops on either side of the cluster. %This may be caused by three potential mechanisms: 
We may rule out (i) inelastic cotunneling processes (higher-order tunneling events) \cite{furusakiTheoryStrongInelastic1995,thielmannCotunnelingCurrentShot2005} such as Franck-Condon coupling to molecular vibrational modes that enable energy exchange during charge transport \cite{leturcqFranckCondonBlockade2009,kochTheoryFranckCondonBlockade2006} (as these peaks should be fixed relative to zero bias rather than to each peak) and (ii) the existence of multiple parallel conduction pathways through distinct molecules or clusters (which would not give the regular spacing observed). It is instead likely that the cluster is sufficiently small that the assumption of very closely spaced energy levels within a metallic dot no longer holds, and the irregular spacing of the quantized levels gives bumps and dips in the density of states.
%If the latter, one can estimate the size from the energy-level spacing of excited states of the nanoparticle, by estimating the average level spacing (the Kubo gap) $\delta \sim \frac{4E_{\rm F}}{3N}$ \cite{Issendorff_Cheshnovsky_2005}, where $\delta$ is the average level spacing, $E_{\rm F}$ is the Fermi energy, and $N$ is the number of conduction electrons. Assuming that the nanoparticle is spherical, the diameter can then be calculated from $d = a \left( \frac{2E_{\rm F} }{ \pi \delta} \right)^{1/3}$, where $a$ is the lattice constant. For a disc-like cluster, its size can be calculated from $d = (\frac{4E_Fa^3}{3 h \pi \delta})^{1/2}$, where $h$ is the height of the disc. 
%$N$ = (Volume of nanoparticle/Volume per atom)×Conduction electrons per atom. For gold, Conduction electrons per atom = 1. For sphere, V = 4/3*pi*R^3. For disc, V = pi*R^2*h. Lattice constant of gold = 0.4078 nm. Volume per atom = a^3/4.

\subsection{Fits to the classical Coulomb-blockade model}\label{secCBfits}

Numerical simulations based on the theory of classical Coulomb blockade \cite{ammanAnalyticSolutionCurrentvoltage1991, hannaVariationCoulombStaircase1991}, where the energy levels form a continuum rather than being discrete, successfully capture all fundamental experimental observations when employing the parameters summarized in Table~\ref{tab:parameters}. Implementation details are provided in \ref{secCBTheory}. Parameters extracted through quantitative fitting include the junction capacitance ratio $C_\text{r} = C_\text{L}/C_\text{R}$, total junction capacitance $C_\text{t} = C_\text{L} + C_\text{R}$, barrier resistance ratio $R_\text{r} = R_\text{L}/R_\text{R}$, and total tunnel resistance $R_\text{t} = R_\text{L} + R_\text{R}$.  Here, $C_\text{L}$ and $C_\text{R}$ are the capacitances to the `left' and `right' electrodes, and $R_\text{L}$, $R_\text{R}$ are the corresponding resistances, respectively, representing asymmetry of the two tunnel barriers.% All of these are consistent with established ranges for metal-island-metal architectures.

The simulated current-voltage characteristics demonstrate excellent quantitative agreement with experimental data in Fig.\ \ref{fig:ivcurves} when three essential physical factors are incorporated: (1) background (residual) charge $Q_0$, which acts as a gate voltage and is constrained to lie in the range $\pm e/2$, (2) broadening phenomena including local heating, inelastic cotunneling processes, and dynamic Coulomb-blockade effects arising from low lead impedance, electrical noise and the applied AC voltage for measuring the conductance with a lock-in amplifier (collectively parameterized through the effective temperature $T$), and (3) the quantum capacitance of the graphene electrode, together with other mechanisms, that affect Fermi-level alignment and so cause a small change in the peak spacing as the voltage goes negative or positive (described by parameters $B$ and $C$ respectively).

The background charge $Q_0$ is critical for reproducing the observed conductance peak position offset with respect to zero bias, with values between $-0.35e$ to $+0.21e$ across all junctions (Table~\ref{tab:parameters}). The variability of this offset likely originates from random static charges in the molecular environment. 

The temperature-dependent broadening can be quantitatively described by the dimensionless temperature parameter $\Gamma_T =  k_\mathrm{B} T/E_\mathrm{c}$ derived for classical Coulomb blockade \cite{beenakkerTheoryCoulombblockadeOscillations1991}. Coulomb blockade dominates when $\Gamma_T\ll 1$. As $\Gamma_T$ increases, the thermal fluctuations gradually wash out the blockade.
%Thermal broadening dominates when $\Gamma_T>1$. 
The fitted $\Gamma_T$ values increase from 0.15 to 0.90 for Junction 1 as the temperature increases from 4 to 128\,K. However, the substantially higher effective temperatures (38--140\,K) extracted from our numerical simulations reveal the presence of additional broadening mechanisms beyond purely thermal effects. As expected, the physical temperature dominates at high temperatures.

\setlength{\tabcolsep}{5pt}
\begin{table}[tb!]
    \centering
    \begin{tabular}{ccccccccc}
        \toprule
        Junction& \(C_{\rm r}\)& \(C_{\rm t}/e\,(V^{-1})\)& \(R_{\rm r}\)& \(R_{\rm t}\,{(\rm M\Omega)}\)& \(Q_0/e\) & \(T\,({\rm K})\)& \(B\) & \(C\)\\
        \midrule
        % Enter your data below. Example row:
        1     & $0.9\scriptstyle\pm0.1$& $17.6\scriptstyle\pm0.9$     & $161\scriptstyle\pm65$  & $9.7\scriptstyle\pm0.1$     & $0.21\scriptstyle\pm0.01$  & $30\scriptstyle\pm7$  & $-0.52\scriptstyle\pm0.04$  & $0.31\scriptstyle\pm0.03$  \\
        $1^\prime$     & 1.2$\scriptstyle\pm0.1$& 14.7$\scriptstyle\pm0.8$& 122$\scriptstyle\pm38$& 9.1$\scriptstyle\pm0.3$& 0.12$\scriptstyle\pm0.01$& 40$\scriptstyle\pm5$& $-0.34\scriptstyle\pm0.08$& 0.18$\scriptstyle\pm0.06$\\
        $1^{\prime\prime}$     & 0.8$\scriptstyle\pm0.1$& 17.9$\scriptstyle\pm0.8$& 109$\scriptstyle\pm20$& 6.5$\scriptstyle\pm0.1$& $-0.26\scriptstyle\pm0.01$& 73$\scriptstyle\pm6$& $-0.56\scriptstyle\pm0.04$& 0.38$\scriptstyle\pm0.03$\\
        9     & $1.1\scriptstyle\pm0.1$     & $14.3\scriptstyle\pm0.4$     & $11.4\scriptstyle\pm1.3$  & $19.6\scriptstyle\pm0.3$     & $-0.21\scriptstyle\pm0.01$  & $24\scriptstyle\pm6$  & $-0.52\scriptstyle\pm0.09$  & $0.10\scriptstyle\pm0.07$\\
        10     & $0.9\scriptstyle\pm0.1$     & $12.6\scriptstyle\pm0.8$     & $8.7\scriptstyle\pm1.6$  & $1.0\scriptstyle\pm0.1$     & $-0.35\scriptstyle\pm0.01$  & $8\scriptstyle\pm7$  & $-0.59\scriptstyle\pm0.16$  & $0.52\scriptstyle\pm0.16$  \\
        11     & $1.3\scriptstyle\pm0.1$     & $9.4\scriptstyle\pm0.2$     & $1.4\scriptstyle\pm0.1$  & $1.2\scriptstyle\pm0.1$     & $0.20\scriptstyle\pm0.01$  & $20\scriptstyle\pm3$  & $0$  & $0$  \\
        12     & $0.7\scriptstyle\pm0.1$     & $5.5\scriptstyle\pm0.4$     & $176\scriptstyle\pm123$  & $686\scriptstyle\pm30$     & $-0.08\scriptstyle\pm0.01$  & $92\scriptstyle\pm19$  & $-0.81\scriptstyle\pm0.16$  & $0.62\scriptstyle\pm0.14$  \\
        13     & $0.9\scriptstyle\pm0.1$     & $5.9\scriptstyle\pm0.4$     & $12.9\scriptstyle\pm2.5$  & $9.8\scriptstyle\pm0.5$     & $0.12\scriptstyle\pm0.01$  & $72\scriptstyle\pm25$  & $-0.53\scriptstyle\pm0.18$  & $1.24\scriptstyle\pm0.27$  \\
        14     & $0.8\scriptstyle\pm0.1$     & $10.7\scriptstyle\pm0.3$     & $8.6\scriptstyle\pm0.8$  & $2.1\scriptstyle\pm0.1$    & $-0.05\scriptstyle\pm0.01$  & $22\scriptstyle\pm7$  & $-0.81\scriptstyle\pm0.18$  & $2.35\scriptstyle\pm0.33$  \\
        % 15     & $1.0\scriptstyle\pm0.1$     & $11.3\scriptstyle\pm0.7$     & $9.9\scriptstyle\pm2.5$  & $2.2\scriptstyle\pm0.1$     & $-0.14\scriptstyle\pm0.01$  & $28\scriptstyle\pm3$  & $-2.11\scriptstyle\pm0.18$  & $0.22\scriptstyle\pm0.21$  \\
        15     & $0.6\scriptstyle\pm0.1$     & $9.1\scriptstyle\pm0.4$     & $237\scriptstyle\pm59$  & $139\scriptstyle\pm6$     & $-0.30\scriptstyle\pm0.01$  & $10\scriptstyle\pm6$  & $-3.60\scriptstyle\pm0.28$  & $2.57\scriptstyle\pm0.23$  \\
        16     & $0.8\scriptstyle\pm0.1$     & $7.8\scriptstyle\pm0.4$     & $17.1\scriptstyle\pm3.6$  & $3.5\scriptstyle\pm0.2$     & $-0.03\scriptstyle\pm0.01$  & $5\scriptstyle\pm1$  & $-0.64\scriptstyle\pm0.11$  & $0$  \\
        \bottomrule
    \end{tabular}
\caption{Table of fitting parameters for each junction: the capacitance ratio $C_\text{r} = C_\text{L}/C_\text{R}$, total junction capacitance $C_\text{t} = C_\text{L} + C_\text{R}$, barrier resistance ratio $R_\text{r} = R_\text{L}/R_\text{R}$, and total tunnel resistance $R_\text{t} = R_\text{L} + R_\text{R}$. % $C_\text{L}$ and $C_\text{R}$ are the capacitances to the `left' and `right' electrodes, and $R_\text{L}$, $R_\text{R}$ are the corresponding resistances, representing asymmetry of the two tunnel barriers.  
$Q_0$ is the background charge (modulo one electron). The fitted temperature $T$ is somewhat higher than the actual temperature (4\,K for all but Junctions $1^{\prime\prime}$, 12 and 13, which were around 77\,K) to account for other forms of broadening not included in the model. $B$ and $C$ are parameters that provide a slight increase in the peak spacing with voltage. The errors are estimates from the fitting.}
\label{tab:parameters}
\end{table}

The quantitative agreement between theory and experiment validates our modified model incorporating graphene-specific corrections. The bias-dependent scaling parameters $B$ and $C$ systematically account for the varying density of states in graphene electrodes, with $-B > C$ in most junctions reflecting, for example, asymmetric band structure near the Dirac point. This asymmetry becomes particularly pronounced in junction 14, suggesting strong electron-hole asymmetry in the local density of states due to doping or strain effects. The variation of tunnel-resistance ratios across all junctions ($R_\text{r} = 8.7$--$237$) indicates substantial differences between nominally identical junctions, probably arising from variations in the distance of the Au cluster from each electrode.
%molecular packing density at electrode interfaces. 
Junction 11's unusually low resistance ratio ($R_\text{r}=1.4\pm0.1$) corresponds to a pair of barriers that are so symmetric that an extra electron is as likely to tunnel into the dot as one is to tunnel out, blurring out all but the first step on either side of zero bias.

The capacitance is related to the dot size. We have modeled the electrostatics of a conducting disk and a sphere, in each case sandwiched between conductors (the gold and graphene electrodes), using the finite-element solver nextnano \cite{birner_nextnano_2007} to obtain the dependence of $C_{\rm t}$ on diameter $d$ (see \extfig Figs.\ \ref{fig:electric_field_norm_sphere} and \ref{fig:electric_field_norm_disc}). The sizes calculated in these ways from each junction's capacitance are given in \extfig Table \ref{tab:capacitance_to_size}.
%The factor of 4 comes from the fact that there are 4 atoms per unit cell in gold, so N = (4/3)*(a^3/4) 
Junction 16 shows a clear subsidiary peak beside the main peak on either side of zero bias, separated by $\approx 42\pm6$ and $38\pm5\,\mathrm{mV}$. The energy is approximately half this, 20\,meV, depending on the junction symmetry. 
%If this is the first excited state, then, from the formula for the Kubo gap above,, gives a diameter of 2.3\,nm for a sphere or $\sim$3.5\,nm for a disk.
%-like cluster, the calculations for standard 2D quantum well give a size $\sim$3.5\,nm.  
The calculated nanoparticle size from the junction capacitance is 3.8\,nm if a sphere, or 3.2\,nm if a thin disk, giving the number of gold atoms as $\sim$75--90. The density of states one can easily calculate for a cuboid with such dimensions has random gaps and peaks, and for this junction a gap may be just above the Fermi energy, for example, whereas in other junctions there is a larger distance to the next peak. 

To investigate the idea that spontaneously formed gold nanoparticles (clusters) are the cause of the staircase characteristics, we fabricated SAMs with the intentional inclusion of a very small proportion of commercially available Au nanoparticles 2.2\,nm or 3.0\,nm in diameter (junctions 12 and 13 in Fig.\ \ref{fig:ivcurves}, respectively). Quite a few of these junctions have low resistance (presumably with many Au nanoparticles in parallel), but others show staircases that are similar to those in pure SAMs, except with a somewhat larger step spacing, such that the fits give total capacitances $C_\text{t}/e = 5.5\pm0.4\,{\rm V}^{-1}$ and $5.9\pm0.4\,{\rm V}^{-1}$, corresponding to spherical nanoparticles with diameter 2.9\,nm and 3.1\,nm.

\subsection{Computational modeling of thiol-Au and amine-Au surface interactions}\label{secModeling}
\begin{figure}[t!]
    \centering
    \includegraphics[width=0.7\linewidth]{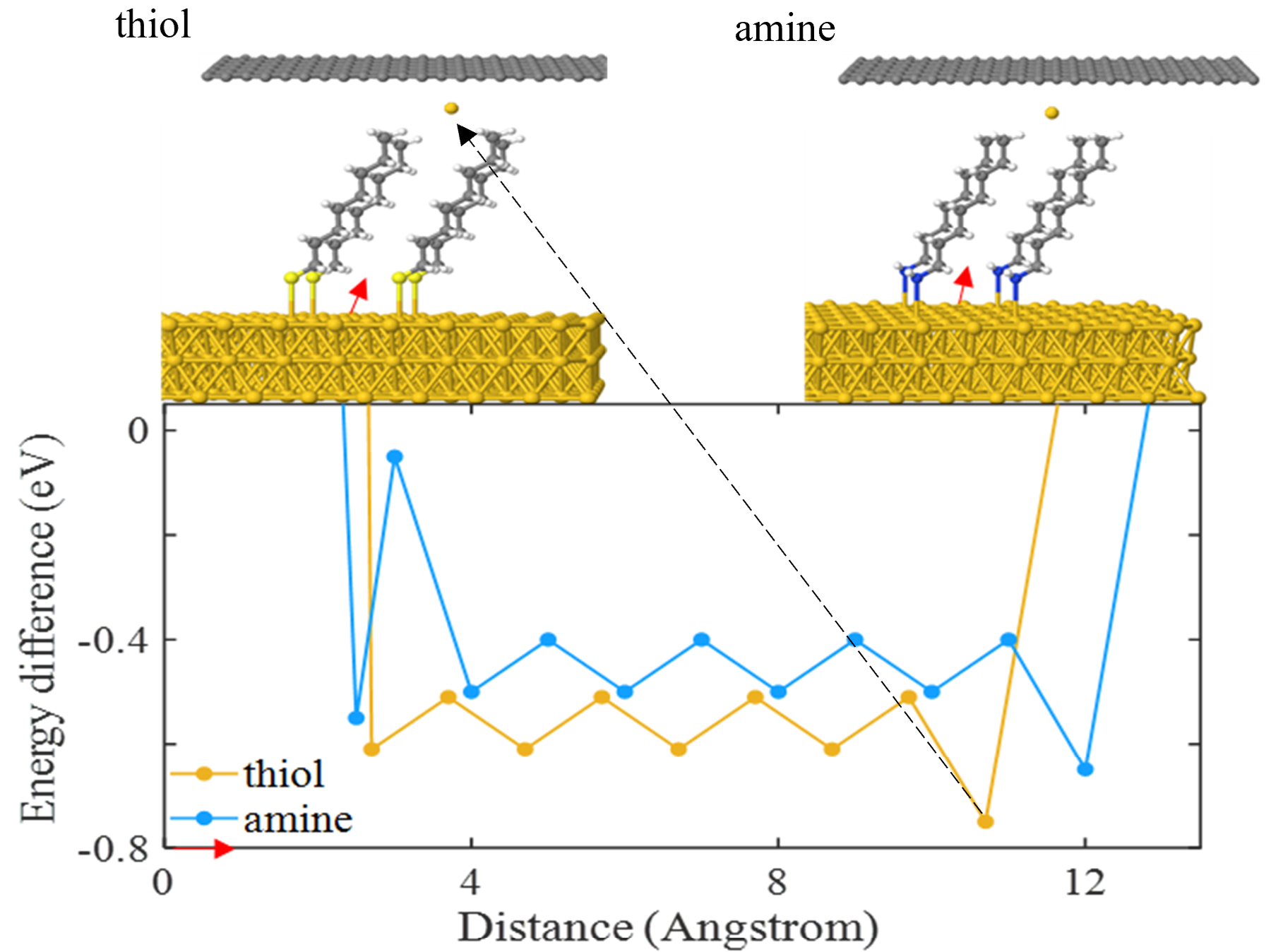}
    \caption{Top panel: Schematics of channels formed from four parallel chains of either thiol- or amine-terminated chains binding to a gold substrate, with a graphene sheet placed on the top. Lower panel: The plots show the variation in the total energy as an Au atom moves along each of the channels (in the direction indicted by the red arrow in the top panel), as a function of the distance from the surface. The yellow and blue plots correspond to thiol- and amine-terminated chains, respectively. }
    \label{fig:theoryMain}
\end{figure}

To investigate the interaction of a gold atom with functionalized alkane chains on a surface, we performed a series of density functional theory (DFT) calculations \cite{soler_siesta_2002,alshehab_impact_2023,wang_enhancing_2024}. Two types of molecules were studied: one terminated with a thiol group (C$_8$S) and the other with an amine group (C$_8$NH$_2$). Each molecule was first geometrically optimized in isolation. They were then individually attached to an Au(111) surface, and the entire molecule-surface system was fully relaxed. Subsequently, a graphene sheet was placed on top of this structure to model the experimental setup \cite{alotaibi_orientational_2024,ismael_three_2024}. The simulation involved moving a single gold atom along the length of the adsorbed chains, starting from the gold surface. More details are given in \ref{secModelingSI}. Fig.\ \ref{fig:theoryMain} shows the resulting variation in the total energy as a function of the Au atom's position.  To manage computational cost, we employed two system sizes: i) For direct comparison of termination effects (thiol vs.\ amine), a smaller system was used, with four shorter chains (C$_8$S, C$_8$NH$_2$) in a periodic arrangement. ii) To analyze the longer-range behavior along a single chain, a larger unit cell was used, containing longer chains (C$_{16}$S, C$_{16}$NH$_2$) (see Supplementary Figs.\ \ref{fig:theoryS4}--\ref{fig:theoryS5}). 

In Supplementary Figs.\ \ref{fig:theoryS4}--\ref{fig:theoryS8}, we show results for the energy profile of an Au atom moving along a single chain, or moving along a channel formed by three, or four, chains in parallel \cite{ismael_energy_2024}. The variation in energy with the position of the Au atom is qualitatively the same in each case. Fig.\ \ref{fig:theoryMain} shows results for a gold atom moving along a channel formed by four chains in parallel. In the case of thiol termination (yellow curve), the first energy minimum occurs at a distance of 2.5\,{\AA}, corresponding to the Au atom being bound to the terminal sulfur. This is followed by an oscillating energy profile, with energy barriers of approximately 0.1\,eV, until a large energy barrier is encountered at the end of the chain, which prevents the Au from penetrating the graphene. In the case of amine termination, a similar energy profile is obtained, except that the first energy barrier at approximately 3\,{\AA} is much larger, taking a value of approximately 0.5\,eV. This energy barrier is 20 times larger than $k_{\rm B} T$ at room temperature, which is sufficient to prevent the Au from leaving the amine group and migrating along the chain. 

It therefore seems likely that molecules anchored with amine groups trap individual gold atoms before they can form clusters, and this is consistent with the lack of Coulomb staircase in such molecules observed here. This may help explain the exceptional reproducibility of amine-anchored molecules reported in break junctions \cite{venkataraman_dependence_2006, quek_aminegold_2007}, where amine anchors bind selectively and weakly to pre-existing under-coordinated Au atoms rather than pulling atoms out of the surface. This preserves the local gold structure, resulting in structural stability and geometry-insensitive conductance, which in turn explains the excellent agreement between experiment and theory in single-molecule junctions. 
Such molecules are also more likely to give consistent behavior in large-area SAMs than those anchored with thiols, and this understanding provides an important pointer to how to improve the yield of molecular electronic devices in the future. More work would be required to determine the behavior when using any other anchors or substrates.

As demonstrated in previous studies \cite{li_uncovering_2009,gao_spontaneous_2018,mom_pressure_2019}, thiol-anchored molecules (RS, where R is the rest of the molecule) can attach to a gold adatom pulled out of the gold surface, or a pair of them can trap an Au adatom between the two sulfurs, in what is called a staple. Even atomically flat Au(111) surfaces can form RS–Au–SR staple motifs because step edges supply a small number of diffusing Au adatoms. Electrolysis of water under a STM tip at a bias of $-1.5$\,V causes formation of many single-monolayer Au islands a few nm across \cite{li_uncovering_2009}, so it is likely that occasional clusters form with other sources of dissociation energy, such as light.
%The formation of these motifs therefore remains an adatom-mediated process, even on nominally flat surfaces. 
These gold atoms are quite unstable, with the Au-SR bond constantly breaking and reforming, and therefore adatoms may be available to form large clusters.
%It is not necessary to invoke the idea that thiol anchors pull atoms out of a gold surface. 
in addition, exposure to ambient air or UV light can oxidize the SAM and weaken or cleave the S–Au bond \cite{kong_influence_2016}. This could lower the activation barrier for gold atom pull-out and promote adatom or staple motif rearrangement. Such structural and charge localization changes might, in turn, lead to the formation of gold clusters.

Atomically flat surfaces, such as made by stripping a layer of gold off a Si template, have a limited availability of adatoms, whereas rougher surfaces, as used in this study as they are scalable, are rich in adatoms and so devices made on them will benefit most from the trapping of the adatoms by the amine group. %It has not been possible yet to make our junctions using template-stripped gold for comparison.

% Hyo Jae Yoon'smessage:

%Additionally, providing surface roughness data (e.g., RMS values or topographical images) would be helpful.

% It is noteworthy that while other molecules such as ferrocene-C11-SH shows no desired charge transport properties on rough surface (e.g., as-deposited surface), BIPY-C11-SH shows desired transport properties regardless of surface roughness. ?Ref?

% Graphene contacts are known to differ significantly from other electrodes, such as EGaIn, by offering low contact resistance and high electronic transparency. I wonder whether these characteristics contribute to the observed phenomena in your study. Related discussion may be added.

\subsection{Gating by tuning background charge}

\begin{figure}[ht!]
    \centering
    \includegraphics[width=\textwidth]{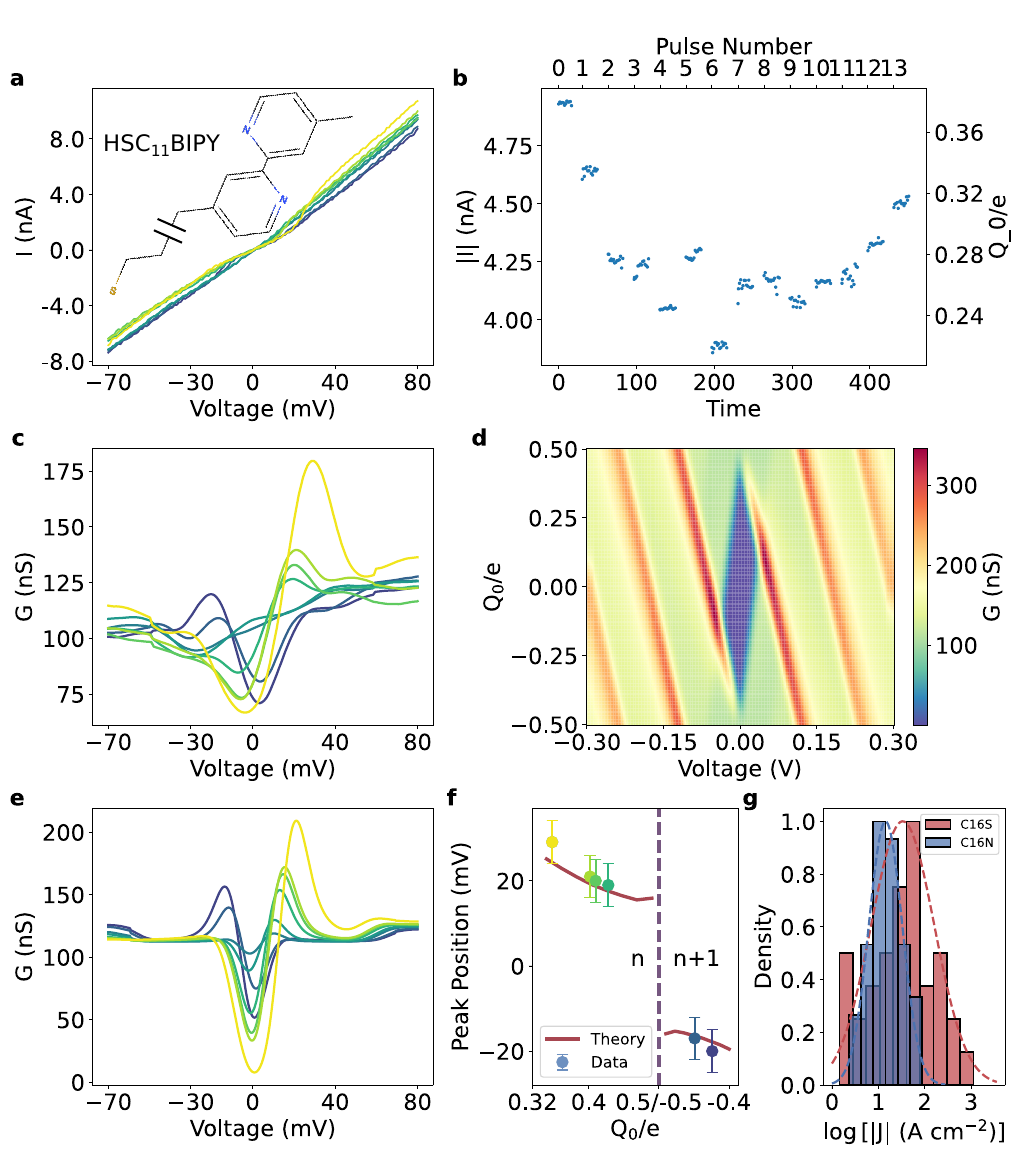}    
    \caption{Background charge gating. (a, c) Measurements of $I$ and $G$ \textit{vs} $V$ with different background charge $Q_0$ (changed somewhat randomly by sweeping to $\pm$0.5\,V), which acts like a gate. These curves were chosen to represent the different $Q_0$ states accessible. Inset in (a): structure of an HSC11BIPY molecule---the pair of lines indicates the omitted central part of the C$_{11}$ backbone. (b) Time-evolution of the current, and hence $Q_0$, for a fixed voltage -50\,mV between each 10\,s voltage pulse (to $-200$\,mV), in a different experiment to that in (a, c). (d, e) Simulated stability diagram and fits to the curves in (c) using $Q_0$ as the only fitting parameter allowed to vary between curves, with the rest of the parameters extracted from the best fits to (a) and (c). (f) Comparison of theoretical and experimental positions of the conductance peaks at various $Q_0$. (g) Histograms of $\mathrm{log_{10} \abs{J}}$ at +0.3\,V for the C16S and C16N junctions, with their Gaussian fits (dashed lines), where $J=I/A$ with $A$ being the nominal junction area.}
    \label{fig:background_charge_gating}
\end{figure}

Tuning the background charge $Q_0$ near the cluster provides a powerful mechanism for controlling charge transport through the junction, effectively gating the cluster by changing the electrostatic potential \cite{brunDynamicalCoulombBlockade2012, schneiderCoulombBlockadePhenomena2013} and hence moving the levels in the dot relative to the chemical potentials in the leads. Figs.~\ref{fig:background_charge_gating}(a) and (c) present experimental $I$--$V$ characteristics and conductance plots for various amounts of background charge, which is consistent with the simulations shown in Fig.~\ref{fig:background_charge_gating}(d),  (e), and (f).
%The shift of the primary peak location from positive to negative bias indicates a change in the net charge $Q = Ne - Q_0$ at which $N$ changes by 1. 
Controlled modulation of $Q_0$ is achieved through voltage pulses, %where the transition probability between charge states 
where the amount of charge injected into the SAM depends critically on both the pulse amplitude and time. Fig.~\ref{fig:background_charge_gating}(b) shows the temporal evolution of both junction current (measured at $-0.05$\,V) and $Q_0$ following each application of a $-0.2$\,V pulse for 10\,s. This result presents the potential for manipulating background charge, as with a gate. However, it is still limited by values of $Q_0$, which are neither continuous nor well-regulated. While the details of the charge injection are unclear, feedback during the pulse could provide a degree of control. Systematic analysis of conductance peak positions across multiple charge states, as illustrated in Fig.~\ref{fig:background_charge_gating}(f), shows quantitative agreement with predictions from the theory of classical single-electron transport (see Section \ref{secCBfits} and \ref{secCBTheory}).

%\subsection{Junction displaying memristive behavior and negative differential resistance}
\subsection{Memristive behavior and negative differential resistance}
\begin{figure}[htb!]
    \centering
    \begin{subfigure}{\textwidth}
    \includegraphics[width=\textwidth]{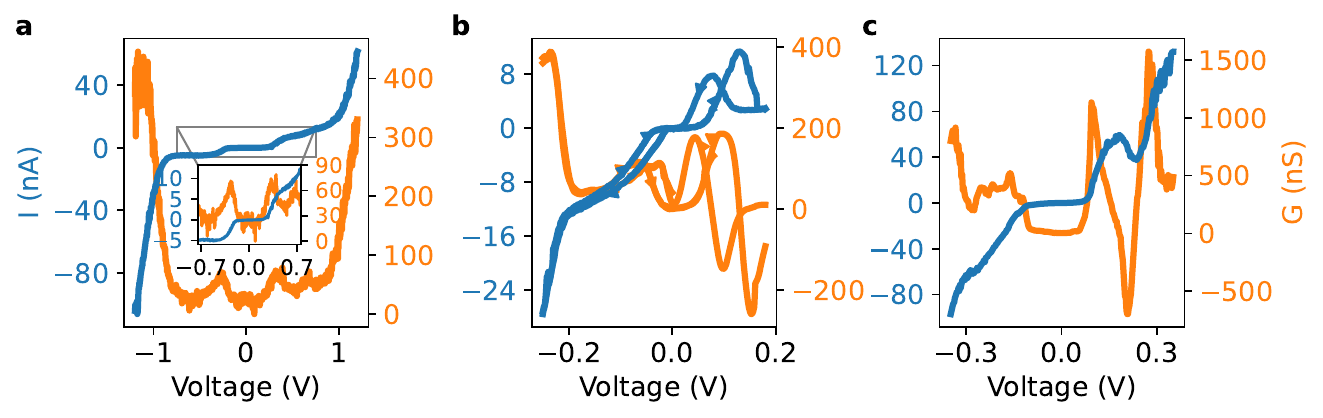}
    \end{subfigure}%
    
    \begin{subfigure}{\textwidth}
    \includegraphics[width=\textwidth]{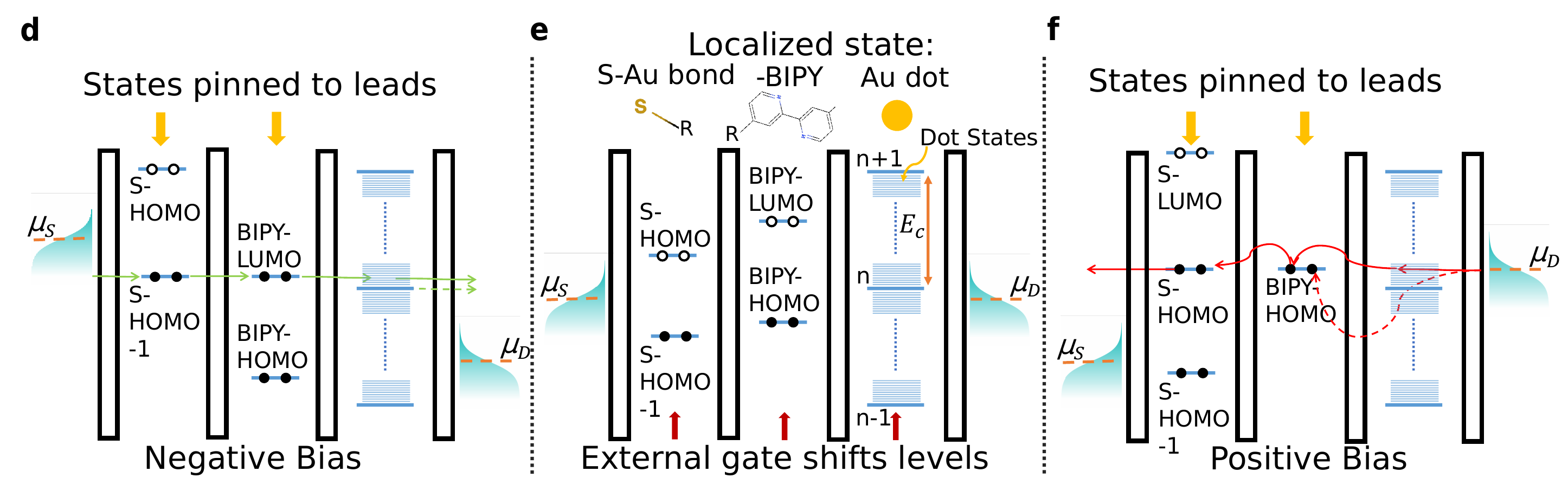}
    \end{subfigure}
    \caption{(a--c): $I$--$V$ (blue) and $G$--$V$ (orange) characteristics of three different junctions. The arrows in (b) indicate the direction in which the voltage is swept. (b, c) simultaneously display Coulomb blockade, NDR, and (in (b)) memristive behavior. (d--f): energy diagrams of a triple-dot system that attempt to explain the mechanism of NDR as the various energy levels come in and out of alignment as the bias is changed.}
    \label{fig:3dot}
\end{figure}

Some HSC11BIPY junctions exhibit additional transport phenomena at cryogenic temperatures, including pronounced current suppression over an extended bias window, rapid current increases at high voltage, and notably, one junction demonstrating memristive switching characteristics, NDR and Coulomb blockade.

Within the $\pm$0.7\,V bias range, Fig.~\ref{fig:3dot}(a) displays characteristic Coulomb-staircase features consistent with a metallic quantum dot at positive bias. However, under negative bias, plateaus are flat beyond the initial conductance step, contrasting with the linear current increase expected from a continuum of electronic states in conventional metallic dots. Beyond this voltage window, a rapid increase in current with applied bias suggests the emergence of additional transport pathways in parallel with metallic-dot conduction. Both Figs.~\ref{fig:3dot}(b) and (c) show distinctive negative-differential conductance (i.e., NDR), and, in addition, Fig.~\ref{fig:3dot}(b) reveals clear memristive behavior (programmability of resistance) together with the NDR through hysteretic $I$--$V$ characteristics, with high-to-low conductance ratios approaching 40:1. The positive-bias regime shows a well-defined NDR region characterized by $G<0$, achieving a peak-to-valley current ratio of around 4:1. After the NDR region, a flat Coulomb-blockade current step emerges, indicating that the energy levels are discrete, unlike metallic quantum dots.

Figs.~\ref{fig:3dot}(d)--(f) present a schematic representation of the proposed triple quantum-dot system. The values of the molecular orbital energies relative to electrode Fermi levels are adopted from previously reported data \cite{parkRectificationMolecularTunneling2021}. Memristive switching results from variation of the background charge $Q_0$, with transitions between two stable configurations. This appears as a shift in the NDR peak position, $\Delta V \approx 0.05$\,V, and a change in the width of the zero-current regime at low bias. The NDR emerges from bias-dependent alignment/misalignment of the localized BIPY-HOMO and S-HOMO energy levels \cite{vanderwielElectronTransportDouble2002,kangInterplayFermiLevel2020,yuanTransitionDirectInverted2018, wulandariMolecularlyanchoredSinglePbS2025}, establishing a voltage-controlled resonant-tunneling condition that changes the transmission probability.

Our measurements reveal two distinct current activation mechanisms. At low bias voltages, charge transport becomes possible only after overcoming the Coulomb charging energy. Subsequent current flow becomes limited by the slowest charge transfer rate in the system. At positive bias, charge transport is predominantly mediated by sequential charge transfer via the drain electrode $\rightarrow$ Au nanoparticle $\rightarrow$ BIPY localized state $\rightarrow$ thiol localized state $\rightarrow$ source-electrode pathway, with the directionality reversed at negative bias. In the linear-response regime of both junctions, charge transport is dominated by conduction through the metallic quantum dot, facilitated by its negligible energy-level spacing. Conversely, within the current-suppressed regime, conduction becomes constrained by tunneling processes involving the BIPY or sulfur-gold localized states, arising from weak electronic coupling between these molecular states and the metallic nanostructure. At high bias voltages where the Fermi level surpasses the HOMO or LUMO, we observe the emergence of additional coherent transport channels through these now energetically accessible molecular levels, yielding a substantial increase in conduction.

\section{Conclusion}\label{concl} 
In summary, we have found evidence for the formation of nanometer-size clusters of gold atoms within a monolayer of molecules self-assembled on a gold surface using a sulfur (thiol) bond on each molecule, with a top graphene electrode. Our data clearly imply that, in thiol-bonded SAMs, isolated gold atoms that were on (or in) the gold electrode cluster together to form islands that can be detected as electrons tunnel through them, provided that the surrounding molecules are so long (e.g., 16 carbon atoms) that conduction through millions of them in parallel is negligible compared to the current through the cluster (`dot'). The two types of tunneling (coherent through the molecules and sequential through the dot) can be distinguished at cryogenic temperatures, as the latter shows Coulomb blockade, owing to the charging energy of adding a single electron to the dot. A current staircase as a function of the voltage $V$ between the gold electrode and a top electrode made of graphene is observed in many samples (junctions). This can be fitted to a Coulomb-blockade model for a dot in which there are many energy levels within the energy window $eV$. Only a metal dot has such closely spaced levels, and the excellent fits show that the dots are metallic and smaller than 7\,nm in diameter, consistent with the formation of a dominant gold cluster.

Thus we have proved that occasional gold clusters form in thiol-anchored SAMs, based on the characteristic shape of a metallic Coulomb staircase, and our other tests provide strong support for this hypothesis. Firstly, deliberately introducing a very small number of 2--3\,nm Au nanoparticles into junctions reproduces the staircases, albeit with slightly smaller capacitance, as would be expected for a sphere rather than a flat disk. Secondly, junctions containing molecules with a different anchor group, namely amine-anchored C16N, show no staircase characteristics. This is consistent with our modelling, which shows that individual Au atoms can be trapped in a potential well next to the amine and so are not free to migrate into the film and form clusters.
%In addition, illumination over many hours, which has been shown to create very large gold clusters.... XPS has found no significant signs of other metals in the SAM...?

In addition, we have shown that the effective gate voltage around the cluster can be tuned by applying a high bias between the electrodes, to move the background charge slightly. This gives an ability to program the conductance of the junction, as would be useful for a memristor. Also, negative differential resistance is sometimes observed, and this may arise from tunneling between states in the two ends of the molecule, and then into a metal cluster.

In future, it is likely that attaching molecules using an amine anchor group ($\mathrm{-NH_2}$) will turn out to produce much more stable and scalable SAMs, and electronic devices, than using a thiol ($\mathrm{-SH}$).

\newpage
\section{Methods}\label{methods}
\subsection{Device fabrication and characterization}

\subsubsection{Fabrication of device substrate}

A 22 mm $\times$ 22 mm silicon chip with a 300\,nm layer of silicon dioxide on top to prevent current leakage to the substrate is patterned to create 36 devices, each containing 40 junctions, resulting in 1440 junctions per batch. The chip is cleaned using acetone, propanol (IPA) and oxygen RF ashing to remove organic contaminants. LOR5B and S1805 photoresists are sequentially spin-coated at 7000\,rpm and 5500\,rpm, respectively, for 60\,s each. LOR5B is baked at 180°C for 10 minutes before S1805 is spun. 
%An edge-bead removal step is also conducted to eliminate accumulation of thick photoresist at the edges, to ensure proper mask-to-wafer contact during exposure. 
The sample is exposed to UV light through a mask for 6.5\,s to define structures followed by 12\,s development in MF319 and nitrogen drying. For the gold contact formation, a 5\,nm titanium layer and a 20\,nm gold layer are deposited by thermal evaporation, followed by lift-off in SVC-14 overnight. For gold connection pads, 20 nm titanium and 80\,nm gold are used, following a similar process. After cleaning with acetone and IPA (5 and 2 minutes, respectively), a 30\,nm AlO$_x$ layer is deposited by electron-beam evaporation as an insulating layer around the junction. After that, a layer of S1805 is spin-coated at 5500\,rpm, UV-exposed for 6.5\,s, developed in MF319 for 12\,s, and the unprotected AlO$_x$ is wet-etched in MF319 for 300 s (etch rate $\sim$ 0.1 nm/s). Finally, the photoresist is removed, and the chips are cleaned in acetone and IPA, ready for self-assembly.

\begin{figure}[b!]
    \centering
    \includegraphics[width=0.8\linewidth]{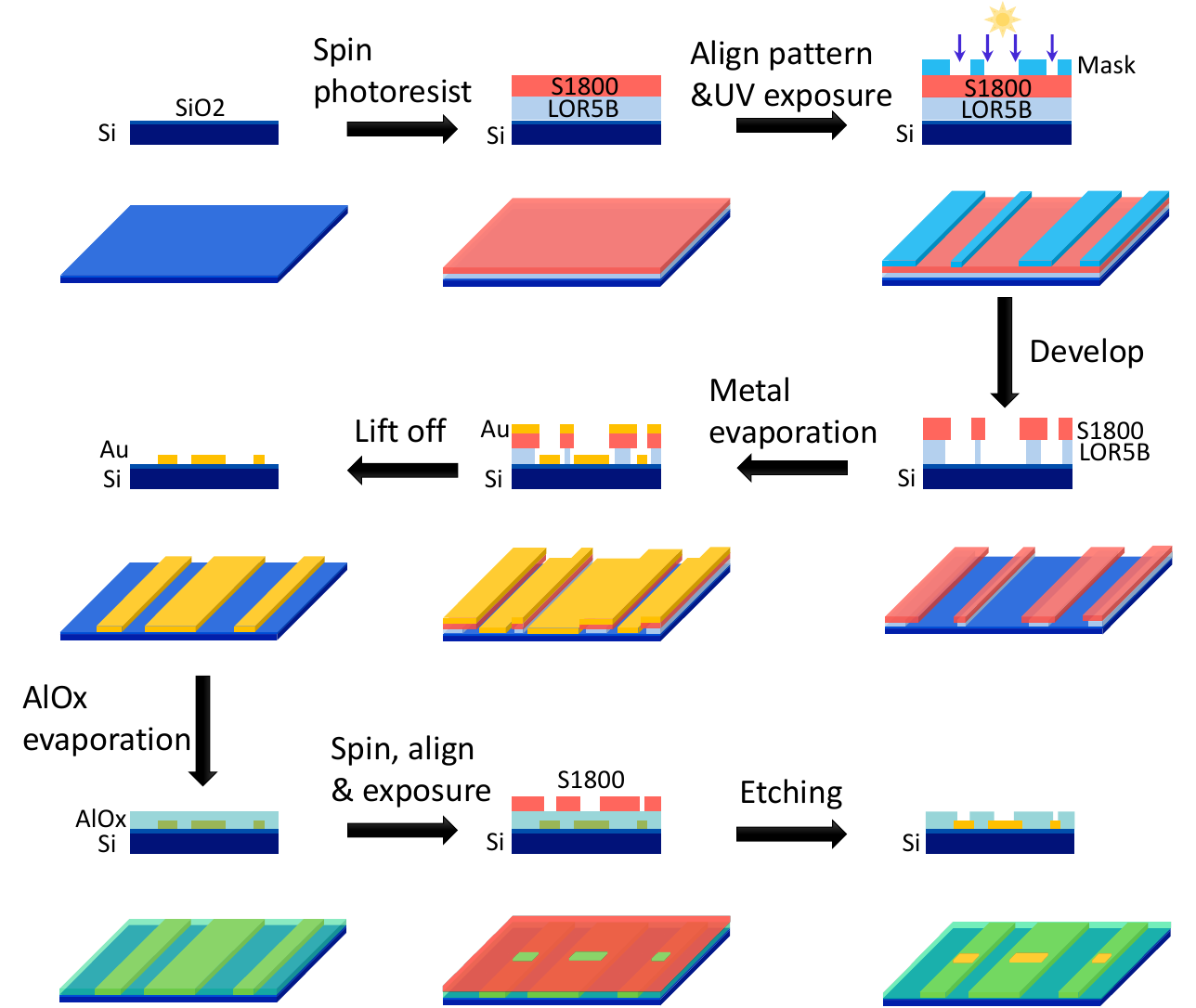}
    \caption{Fabrication process.}
    \label{fig:fabrication}
\end{figure}

\subsubsection{Standard self-assembly}

For alkanethiols or alkanedithiols (purchased from Sigma-Aldrich), the molecules are dissolved in DMF with 1\,mM concentration. The device, after cleaning and RF oxygen plasma ashing, was immersed in the solution for 25 minutes, washed with ethanol (EtOH) twice followed by DMF, and re-immersed in the solution. This process was repeated 5 or 6 times, and after that, the device was immersed in the solution for 24--48 hours for SAM self-assembly. XPS experiments were performed to verify the growth of the SAM.

\subsubsection{Repeated surface exchange of molecules (ReSEM)}

We have fabricated and measured molecular junctions with a SAM of 2,2'-bipyridyl-terminated undecanethiol ($\mathrm{HSC_{11}BIPY}$) and $n$-alkane thiol (HSC$_{n}$), where $n=12,14,16,18$. These have large head groups, which limit the packing density, and we have shown that SAM quality, and hence the breakdown voltage, can be greatly improved by assembling smaller molecules in any remaining gaps in the SAM, by using the ReSEM technique. Here, alkanethiols in the form of HS\text{-}(CH$_2$)$_{n}$-SH (`C$n$S2') with $n=6,8,10$ or $\mathrm{(CH_2)_{8}\text{-}SH}$  (`C$n$S') were used as the supporting molecules \cite{kongInterstitiallyMixedSelfAssembled2021,jangElectricallyStableSelfAssembled2024}. There is no clear correlation between supporting molecules employed and the cluster size or the yield of the junctions displaying Coulomb staircase. %M recipe are described in \nameref{methods}. 

The ReSEM technique is different from our standard self-assembly. First of all, 5\,ml of 1mM HSC11BIPY and $\mathrm{HS\text{-}(CH_2)_{n}\text{-}SH}$ (purchased from Sigma-Aldrich) are prepared, following our established recipe for molecules with large head groups \cite{kongInterstitiallyMixedSelfAssembled2021}. Following cleaning and processing in an oxygen RF asher, the substrate is then immersed in HSC11BIPY for 3 hours. The device is then immersed in the supporting molecule solution and HSC11BIPY for 3 hours and 18 hours, respectively. The previous step is repeated twice more, followed by cleaning in ethanol, acetone and IPA. ReSEM typically achieves higher molecular density than corresponding single-component SAMs, and longer alkyl chains tend to exhibit better molecular packing than shorter ones.

\subsubsection{Graphene transfer}

Single-layer graphene, grown by chemical vapor deposition (CVD) on copper (ACS Material) was used in this work. The copper etching procedure followed the standard recipe for graphene transfer, where 5\,g ammonium persulfate (APS) is dissolved in 150\,ml water as the etchant solution. PMMA is spun on the CVD graphene surface and the graphene is cut to the desired dimensions before etching. After etching in the APS solution overnight, the graphene is transferred into clean DI water 6 times to wash off the remaining APS, and the PMMA-protected graphene is eventually transferred onto the chip with a SAM using the common ‘fishing’ technique.

The device after graphene transfer was incubated in a high vacuum (10$^{-7}$\,mbar) overnight to remove residual water. After this drying stage, the PMMA was washed off by immersing the device in acetone for 24 hours, followed by cleaning in acetone and IPA. Photolithography was conducted to pattern and protect the graphene electrodes. The unprotected region was then removed by oxygen RF ashing at 12\,W for 240\,s. 

The typical graphene quality and electrical contact integrity were systematically investigated using complementary characterization techniques. Raman spectroscopy of a representative piece of graphene showed a G/D peak ratio exceeding 4.5 with the 2D peak's FWHM below \textless 35\,cm$^{-1}$, see \extfig Fig. \ref{fig:raman}. This indicates high-quality graphene with minimal defects~\cite{ferrariRamanSpectroscopyVersatile2013}. Ohmic contact behavior and interface stability were validated through control devices fabricated without molecular layers, demonstrating Au-graphene contact resistances consistently below 2\,k$\Omega$ across 40 junctions.

\subsubsection{Au nanoparticle deposition}

To deliberately introduce metallic nanoparticle contamination into the junction, we first prepare the Au nanoparticles (2.2\,nm or 3\,nm in diameter, coated in PVP, from NanoPartz) dissolved in ethanol with the desired concentration. For the data shown in this paper, 0.6\,ng/mL were adopted. Before any self-assembly of molecules, the substrate is immersed in the nanoparticle solution for 5\,min, followed by cleaning in EtOH.

\subsubsection{Electrical measurements}

Electrical transport measurements were conducted using a liquid-helium or liquid-nitrogen dip station % (4.2--300\,K)  
and an Oxford Instruments liquid-helium cryostat with a variable-temperature insert and a superconducting magnet. % (2--300\,K). 
The $I$--$V$ characteristics were measured using Keithley SMUs (236, 2400, 2635) and the conductance $G$ was measured with a Signal Recovery 7265 lock-in amplifier with AC of 1 or 10\,mV rms amplitude at 189.3\,Hz added on to the DC from the SMU by floating the SMU ground on the AC signal. The microwell array architecture enables parallel electrical measurement of arbitrarily many junctions per chip (20 or 40 in our case), with individual active areas defined precisely through lithographically patterned AlO$_x$ apertures with areas ranging from 2.0\,$\pm$\,1.0 to 14.0\,$\pm$\,2.0\,$\upmu$m$^2$. Systematic analysis revealed that junctions exhibiting both contact resistances exceeding 10\,k\ohm, and current-voltage characteristics that were nonlinear demonstrated a yield \textgreater 80\%, with performance variations correlating strongly with molecules and self-assembly protocols.

\subsubsection{X-ray photoelectron spectroscopy}

XPS was performed using a Kratos Analytical AXIS Supra spectrometer with monochromatic Al K$\alpha$ 1486.7\,eV X-ray source, operating at 15\,kV, 15\,mA. All the spectra were analyzed using CASAXPS (Casa Software Ltd, UK).

\subsubsection{Topographical imaging}

Graphene film morphology was characterized using a Bruker MultiMode 8 atomic force microscope (AFM) with Nanoscope 6 controller in PeakForce tapping mode (force set point 10\,nN) using a Spark 70\,Pt (Nu Nano Ltd) AFM probe (spring constant $k = 2.0$\,N/m).

\subsubsection{Ultrasonic Force Microscopy}

Ultrasonic Force Microscopy (UFM) is a modification of standard contact-mode AFM in which a high-frequency (4.11\,MHz) vertical oscillation is applied to the sample via a dedicated piezoceramic transducer. Amplitude modulation (2.7\,kHz, sawtooth) is applied to this high-frequency drive signal, and the amplitude of the cantilever deflection at the modulation frequency is detected using a lock-in amplifier (SR830, Stanford Research Systems). The resulting signal corresponds to the nonlinear UFM response, which is highly dependent on the local nanomechanical properties of the sample, and has been shown to effective in mapping local stiffness in supported and suspended graphene films \cite{kay_electromechanical_2014, mucientes_mapping_2020}. UFM measurements were carried out using an OPUS 240AC-NA (MikroMasch) probe.

\bibliography{sn-bibliography}% common bib file
%% if required, the content of .bbl file can be included here once bbl is generated
%%\input sn-article.bbl

\bmhead{Acknowledgements}

This work was supported by the UK EPSRC (grants EP/P027156/1, EP/P027172/1 and EP/P027520/1) and the UKRI programme grant EP/X026876/1 `QMol'. Synthesis of the BIPY molecules was supported by the National Research Foundation (NRF) of Korea (RS-2024-00351264) and a Korea University Grant. Y.L. acknowledges funding from Girton College, Cambridge, UK.

\section*{Data availability}

All data supporting the findings of this study are presented in the Article and its Supplementary Information. Source data are provided with this paper.

\section*{Code availability}

The code for the Coulomb-blockade simulations is available at [University of Cambridge repository...].

\section*{Author Information}

\noindent\textbf{Shanglong Ning}

\noindent Present address: Blackett Laboratory, Imperial College London, South Kensington Campus, London, SW7 2AZ, United Kingdom.

\noindent\textbf{Xintai Wang}

\noindent Present address: College of Environmental Science and Engineering, Dalian Maritime University, Dalian, China, and Zhejiang Mashang Technology Research Institute, Cangnan, Wenzhou, Zhejiang, China.

%\subsection*{Authors and Affiliations}
%\textbf{University of Cambridge, ...}\\
%Bingxin Li, Chunyang Miao, ...., Christopher Ford

\subsection*{Contributions}

%C.J.B.F. conceived and supervised the project, S.N., B.L., C.M., C.J.B.F. designed the experiments,  S.N., B.L., C.M., V.C., Y.L. performed all the transport measurements and S.N., B.L., C.M. analyzed the data. S.N. developed the initial simulation framework in MATLAB; B.L. improved the theory and implemented the code in Python and performed all simulations. X.W. and S.N. developed the alkanethiol SAM growth protocol. X.W. performed initial C16S room-temperature measurement. H.J.Y. supervised the project to synthesize and characterize the BIPY molecule and develop the ReSEM method. G.D.K. synthesized the BIPY molecule and conducted experiments to optimize the ReSEM conditions. C.G. carried out the dark-field imaging and exposure tests with direction from J.J.B.. S.J., J.N. and B.R. performed the XPS measurements. S.H., O.K.\ and B.R. performed the AFM and UFM measurements. B.L., S.N., C.M., and C.J.B.F. wrote the manuscript with input from all authors.

CJBF conceived and supervised the project, SN, BL, CM, CJBF designed the experiments,  SN, BL, CM, VC, YL performed all the transport measurements and SN, BL, CM analyzed the data. SN developed the initial simulation framework in MATLAB; BL improved the theory and implemented the code in Python and performed all simulations. XW and SN developed the alkanethiol SAM growth protocol. XW performed initial C16S room-temperature measurement. HJY supervised the project to synthesize and characterize the BIPY molecule and develop the ReSEM method. GDK synthesized the BIPY molecule and conducted experiments to optimize the ReSEM conditions. CG carried out the dark-field imaging and exposure tests with direction from JJB. SJ, JN and BR performed the XPS measurements. SH, OK and BR performed the AFM and UFM measurements. BL, SN, CM, and CJBF wrote the manuscript with input from all authors.

%Hyo Jae Yoon: supervised the BIPY molecule-related project, including the synthesis and characterization of the molecule, development of the ReSEM method, and initiated this collaboration following a conference discussion with Prof. Chris Ford.

%Gyu Don Kong: synthesized the BIPY molecule and conducted experiments for optimizing the ReSEM conditions.

\subsection{Corresponding author}
Correspondence to Bingxin Li, bl497@cantab.ac.uk;
Shanglong Ning, sn538@cantab.ac.uk;
Chunyang Miao, cm2109@cam.ac.uk;
Christopher Ford, cjbf@cam.ac.uk.

\section{Ethics declarations}
\subsection{Competing interests}

The authors declare no competing interests.

%\section{Peer review}
%\subsection{Peer review information}

%Nature Nanotechnology thanks the anonymous reviewer(s) for their contribution to the peer review of this work.

%\section{Additional information}

%\textbf{Publisher’s note} Springer Nature remains neutral with regard to jurisdictional claims in published maps and institutional affiliations.

\clearpage

\backmatter
\renewcommand{\figurename}{\extfig Figure}
\setcounter{figure}{0}
\renewcommand{\tablename}{\extfig Table}
\setcounter{table}{0}
%\bmhead{Extended information}
\section*{Extended Data}

\begin{figure}[h!]
    \centering
    \includegraphics[width=0.4\linewidth]{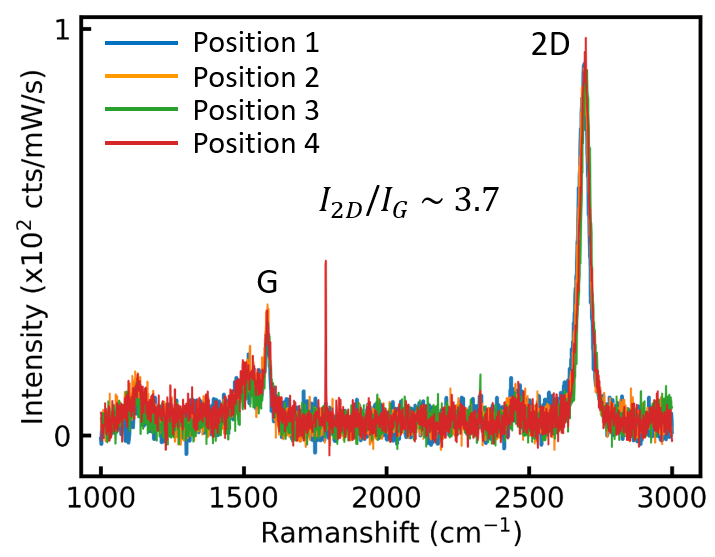}
    \caption{Raman Spectroscopy of a representative piece of graphene on a device.}
    \label{fig:raman}
\end{figure}

\begin{figure}[h!]
    \centering
    \includegraphics[width=0.7\linewidth]{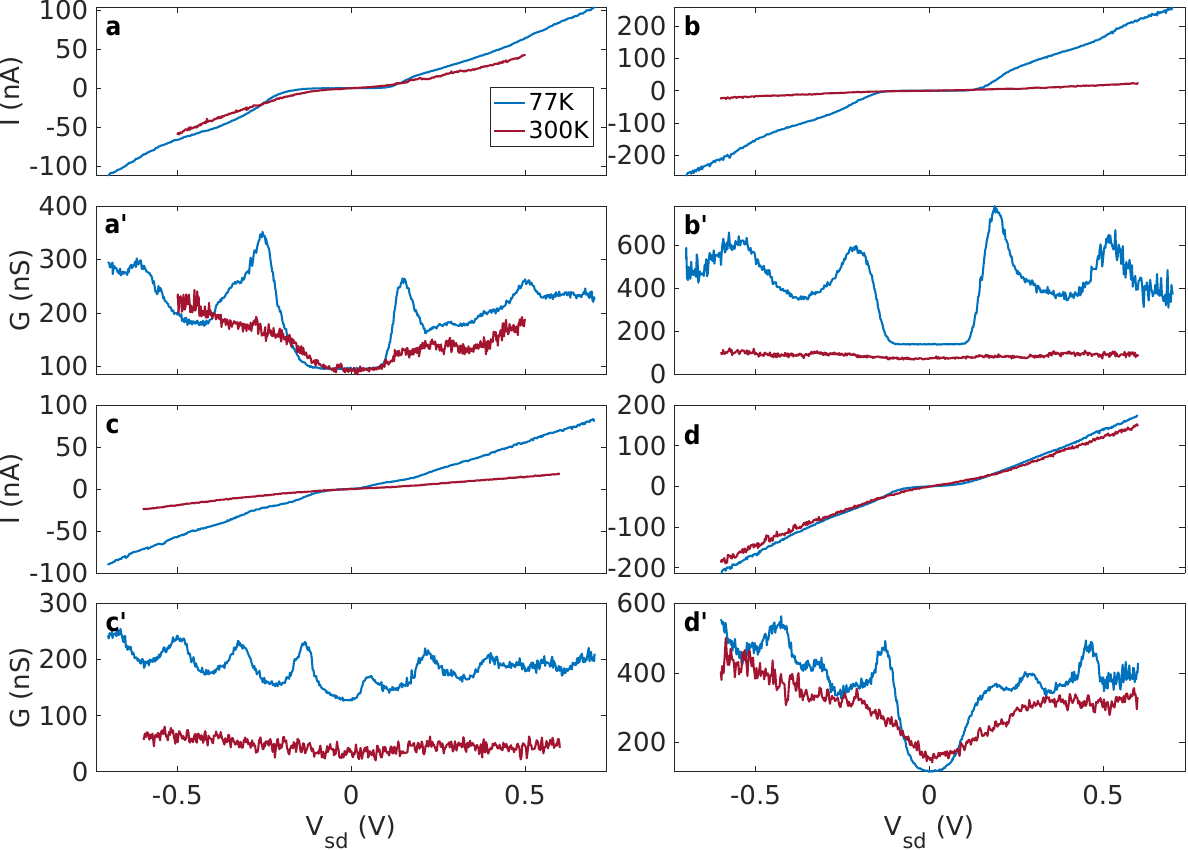}
    \caption{Comparison of the I--V and G--V curves for four example junctions displaying Coulomb staircase at 77\,K (blue) and 300\,K (red). The data at 300\,K were taken before low-temperature measurements. These are all HSC11BIPY junctions.}
    \label{fig:temperature compare}
\end{figure}

\begin{figure}[h!]
    \centering
    \includegraphics[width=\linewidth]{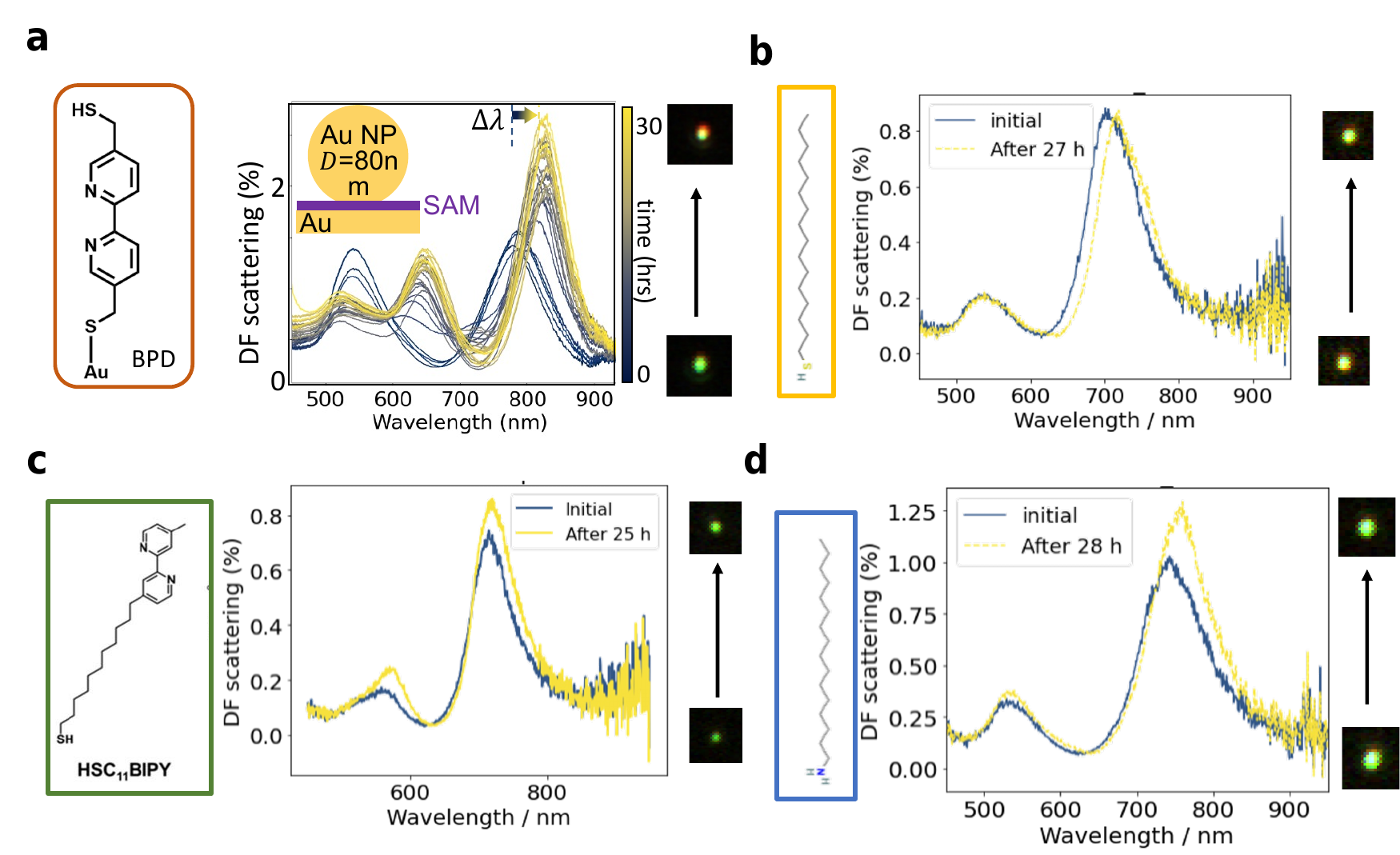}
    \caption{Dark-field (DF) scattering spectra in a nanoparticle-on-mirror (NPoM) gap under 0.1\,$\mathrm{W/cm^2}$ visible illumination. The different molecules used, as shown to the left of each panel, were (a) BPD \cite{guoExtensivePhotochemicalRestructuring2024}, (b) C16S, (c) HSC11BIPY, and (d) C16N. A peak shift indicates a change in the facet of the NPoM gap or the size of the gap. (a) shows a clear peak shift, while (c) shows very little shift after illumination. There is no noticeable shift in the peaks in (b) and (d), indicating no detectable changes in the nanoparticle's facet.
    % (a) is reproduced with permission from \cite{guoExtensivePhotochemicalRestructuring2024}.
    }
    \label{fig:optics} % not referred to
\end{figure}

\begin{figure}[h!]
    \centering
    \includegraphics[width=\linewidth]{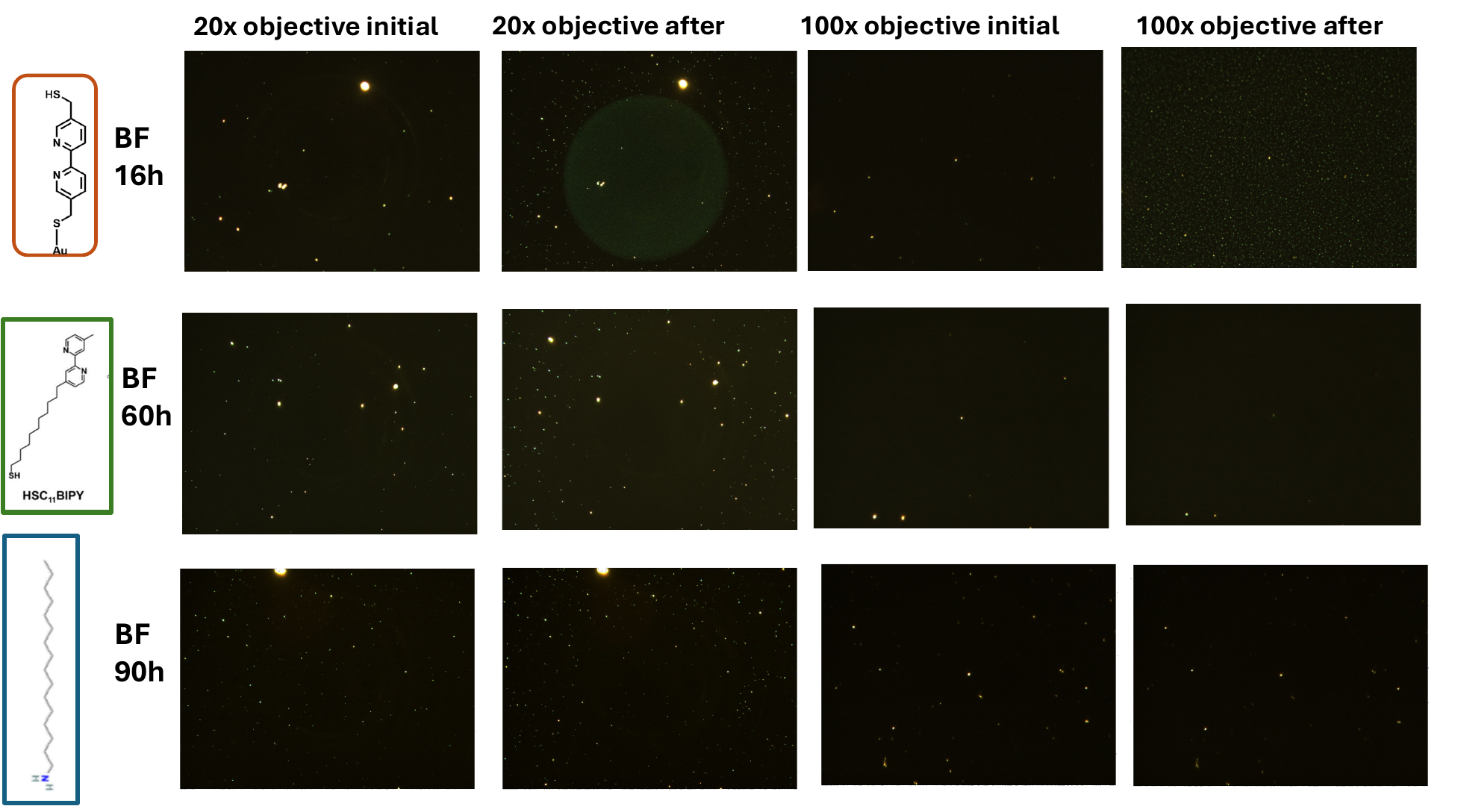}
    \caption{Time-evolution of different SAM on template stripped gold under white light exposure. BPD exhibits distinct visible gold clusters in the illuminated area, while the other two molecules show minimal changes.}
    \label{fig:BF illumination}
\end{figure}

\begin{figure}[h!]
    \centering
    \includegraphics[width=0.8\linewidth]{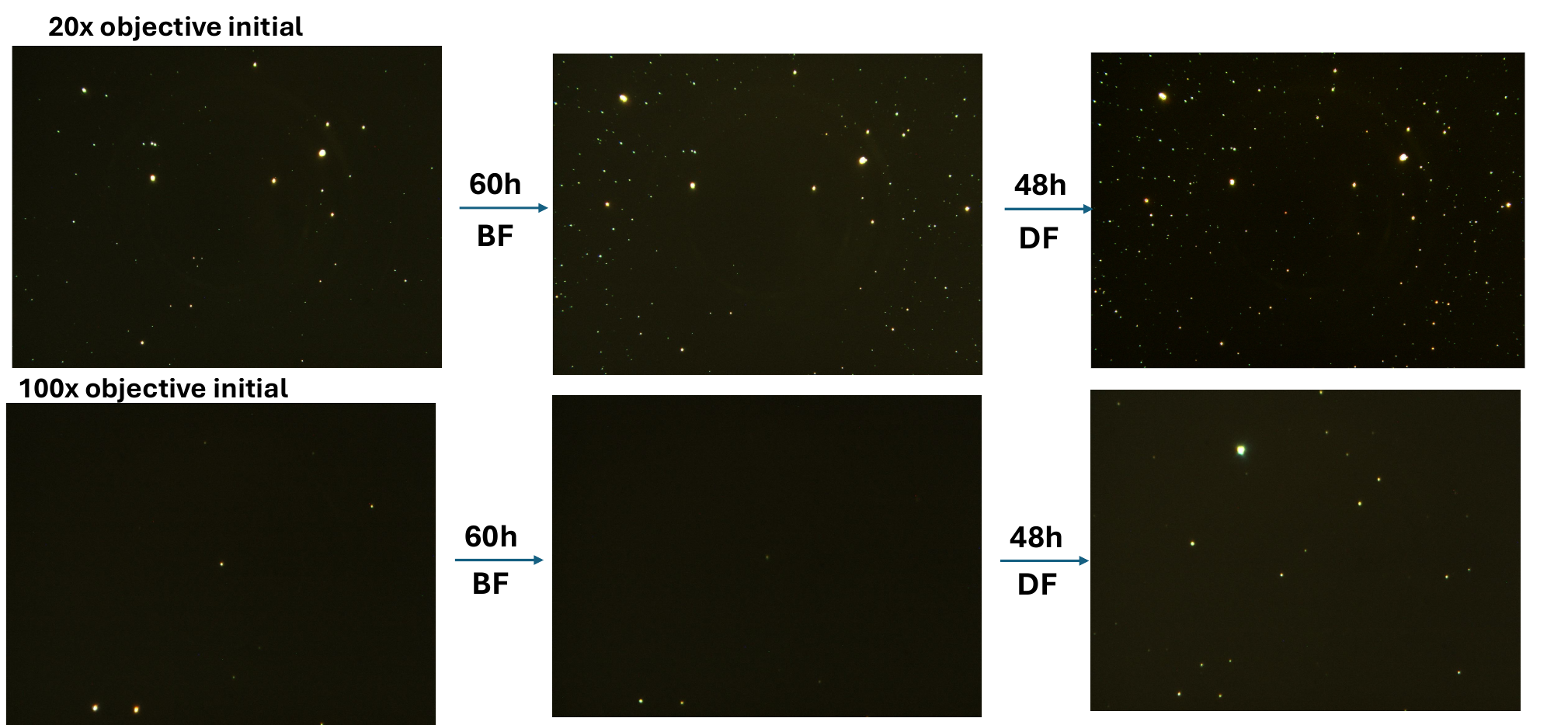}
    \caption{Time-evolution of HSC11BIPY SAM on template-stripped gold under bright field white light and dark filed white light. No visible gold clusters form under bright-field (BF) illumination, where light is perpendicular to the gold surface. However, some clusters do form under dark-field (DF) illumination, where light shines at an angle. This suggests that the activation and migration of gold nanoparticles may depend on whether light can reach the gold surface.}
    \label{fig:DF vs BF}
\end{figure}

\begin{figure}[t!]
    \centering
    \begin{subfigure}{0.45\textwidth}
        \includegraphics[width=\textwidth]{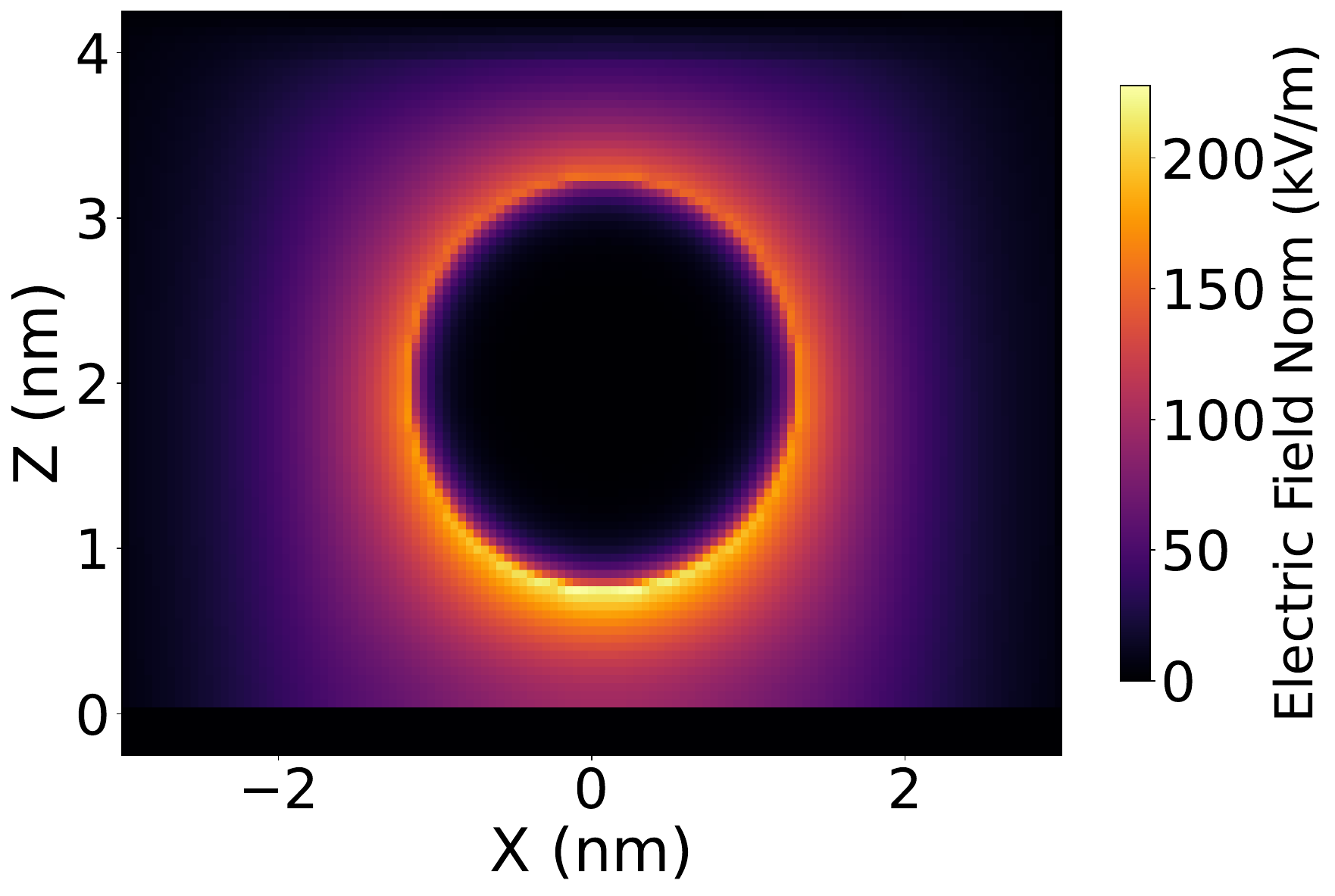}
        \caption{}
    \end{subfigure}
    \begin{subfigure}{0.3\textwidth}
        \includegraphics[width=\textwidth]{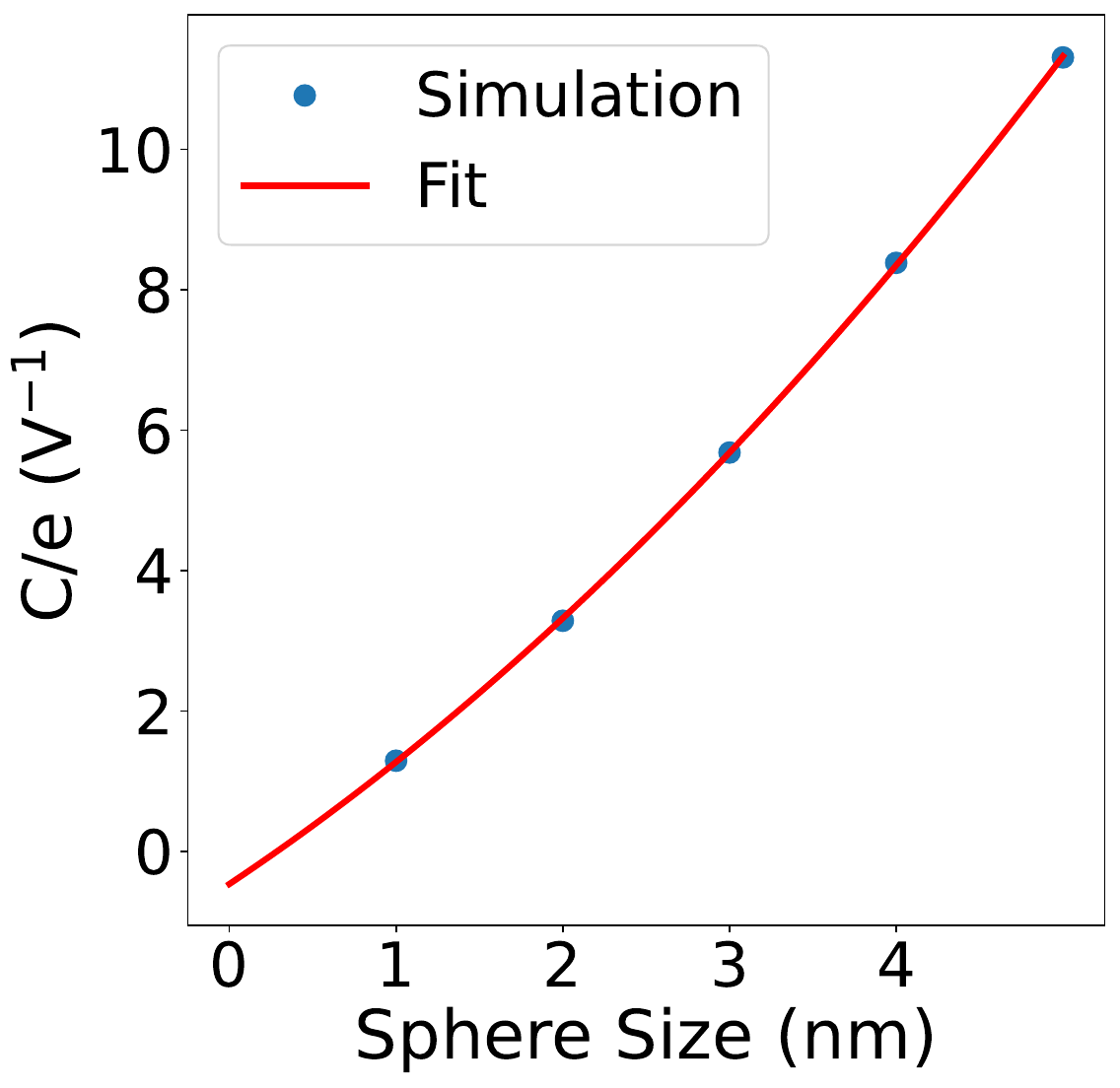}
        \caption{}
    \end{subfigure}
    \caption[Simulation of spherical Au cluster]{(a) Example of electrostatic simulation of the differential capacitance of a gold sphere sandwiched between two metal electrodes. The gaps between the sphere and the electrodes are set to 0.5\,nm. The relative dielectric constant $\upepsilon_r$ is set to 3, an estimate of the SAM value. (b) Simulation and fitted curve ($C/e = 0.16d^2 + 1.58d - 0.46$, where $d$ is the size of the sphere) for $C/e$ vs sphere size.}
    \label{fig:electric_field_norm_sphere}
\end{figure}

\begin{figure}[t!]
    \centering
    \begin{subfigure}{0.45\textwidth}
        \includegraphics[width=\textwidth]{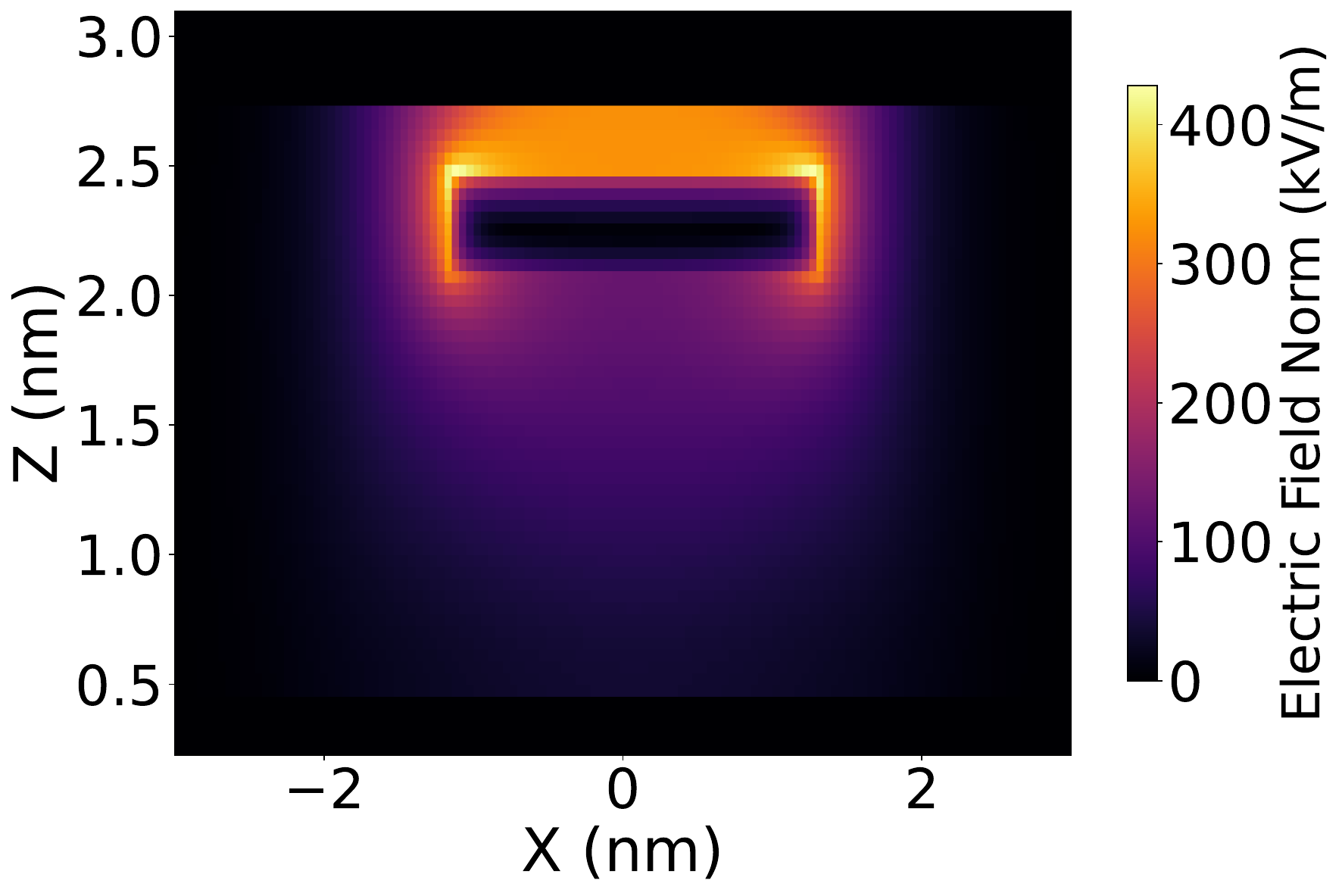}
        \caption{}
    \end{subfigure}
    \begin{subfigure}{0.3\textwidth}
        \includegraphics[width=\textwidth]{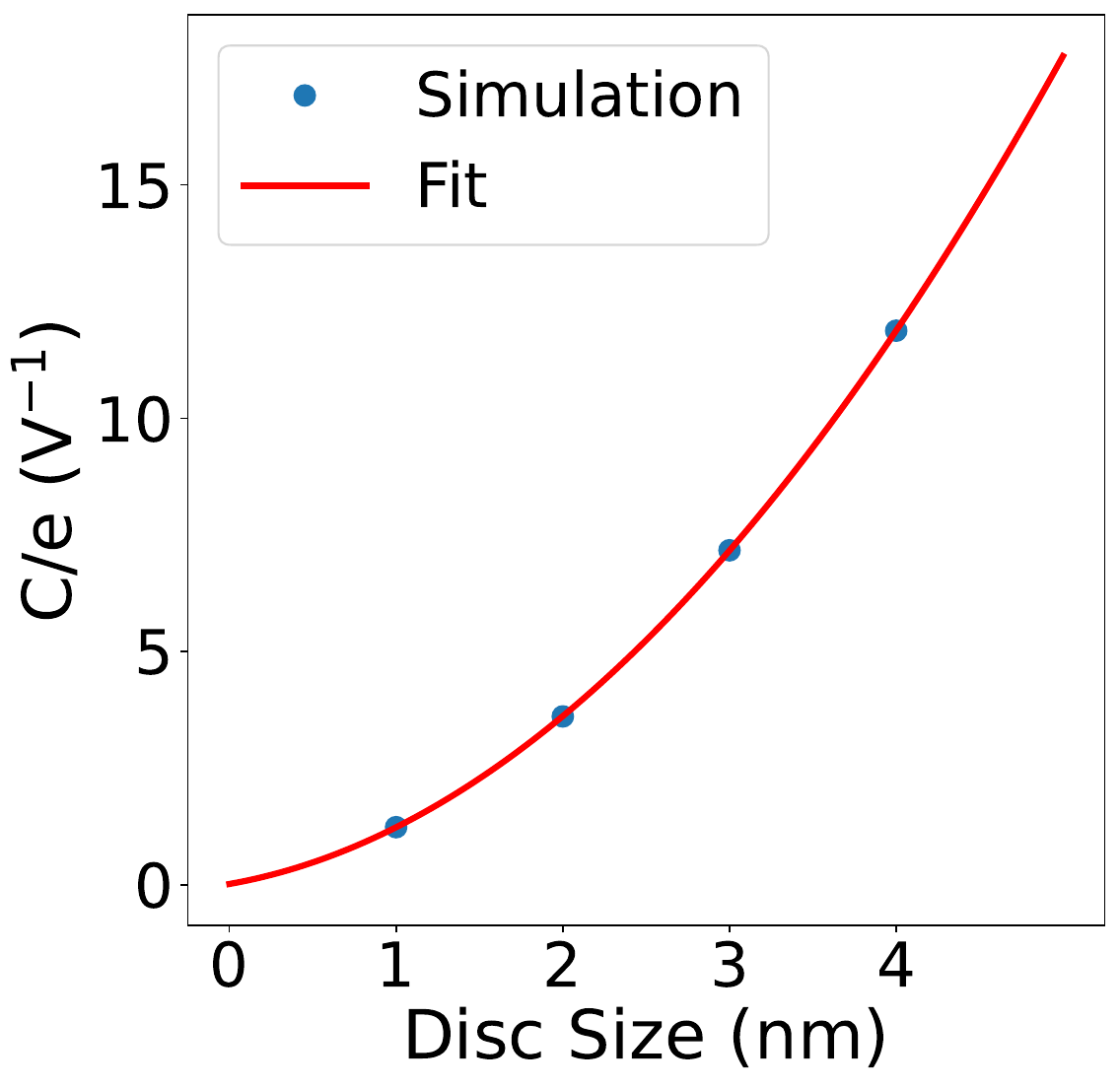}
        \caption{}
    \end{subfigure}
    \caption[Simulation of a disk-shaped Au cluster]{(a) Example of electrostatic simulation of the differential capacitance of a gold disk 0.3\,nm thick (coarsely approximating one layer of atoms) sandwiched between two metal electrodes. The gaps are set to 0.2\,nm and 1.0\,nm for the top and bottom electrodes, respectively. The relative dielectric constant $\upepsilon_r$ is set to 3, an estimate of the SAM value. (b) Simulation and fitted curve ($C/e = 0.58d^2 + 0.63d + 0.02$, where $d$ is the size of the disk) for $C/e$ vs disk size.}
    \label{fig:electric_field_norm_disc}
\end{figure}
\FloatBarrier

\setlength{\tabcolsep}{10pt}
\begin{table}[t!]
    \centering
    \begin{tabular}{cccc}
        \toprule
        Junction& $C_{\rm t}/e\rm{(V^{-1})}$ & Disk Diameter (nm)& Sphere Size (nm)\\
        \midrule
        1          & 17.6$\scriptstyle\pm0.9$   & 5.0$\scriptstyle\pm0.3$ & 6.8$\scriptstyle\pm0.3$ \\
        9          & 14.3$\scriptstyle\pm0.4$   & 4.4$\scriptstyle\pm0.1$ & 5.9$\scriptstyle\pm0.2$ \\
        10          & 12.6$\scriptstyle\pm0.8$   & 4.1$\scriptstyle\pm0.3$ & 5.4$\scriptstyle\pm0.3$ \\
        11          & 9.4$\scriptstyle\pm0.2$    & 3.5$\scriptstyle\pm0.1$ & 4.4$\scriptstyle\pm0.1$ \\
        12          & 5.5$\scriptstyle\pm0.4$    & 2.6$\scriptstyle\pm0.2$ & 2.9$\scriptstyle\pm0.2$ \\
        13          & 5.9$\scriptstyle\pm0.4$    & 2.7$\scriptstyle\pm0.2$ & 3.1$\scriptstyle\pm0.2$ \\
        14          & 10.7$\scriptstyle\pm0.3$   & 3.8$\scriptstyle\pm0.1$ & 4.8$\scriptstyle\pm0.1$ \\
        15          & 9.1$\scriptstyle\pm0.4$    & 3.4$\scriptstyle\pm0.1$ & 4.3$\scriptstyle\pm0.2$ \\
        16          & 7.8$\scriptstyle\pm0.4$    & 3.2$\scriptstyle\pm0.2$ & 3.8$\scriptstyle\pm0.2$ \\
        \bottomrule
    \end{tabular}
    \caption[Fitted cluster sizes]{Fitting cluster size from the capacitance. The sphere and disk diameters are interpolated from the corresponding simulated capacitance values shown in \extfig Figs.\ \ref{fig:electric_field_norm_sphere}(b) and \ref{fig:electric_field_norm_disc}(b). These are just estimates of the sizes, given the large uncertainty in the cluster shape and the dielectric constant of the SAM.
    %disc size is calculated by setting the disc height to 0.3\,nm (one layer of atoms) with a 0.2\,nm gap to the top electrode and 1\,nm gap to the bottom electrode. The sphere size is calculated with a 0.5\,nm gap to the top and bottom electrodes.
    }
    \label{tab:capacitance_to_size}
\end{table}

\end{document}